\documentstyle[12pt]{article} 
\def\lsim{\mathrel{\lower2.5pt\vbox{\lineskip=0pt\baselineskip=0pt 
           \hbox{$<$}\hbox{$\sim$}}}} 
\def\gsim{\mathrel{\lower2.5pt\vbox{\lineskip=0pt\baselineskip=0pt 
           \hbox{$>$}\hbox{$\sim$}}}} 
\renewcommand{\theequation}{\thesection.\arabic{equation}}
           
\begin{document}
\setlength{\baselineskip}{8mm}
\setcounter{page}{1}
\thispagestyle{empty}
\begin{flushright}
\ DPNU-96-23, \  AUE-96/02 \\
May 1996 \\
\end{flushright}
\vspace{5mm}
\begin{center}
{\large  \bf 
Fermion Masses and Mixings \\
in a String Inspired Model }
\end{center}
\vskip 10mm
\begin{center} 
Naoyuki HABA$^1$, Chuichiro HATTORI$^2$, \\
Masahisa MATSUDA$^3$ and Takeo MATSUOKA$^1$  \\
{\it 
${}^1$Department of Physics, Nagoya University \\
           Nagoya, JAPAN 464-01 \\
${}^2$Science Division, General Education \\
     Aichi Institute of Technology \\
      Toyota, Aichi, JAPAN 470-03 \\
${}^3$Department of Physics and Astronomy \\
     Aichi University of Education \\
      Kariya, Aichi, JAPAN 448 \\
}
\end{center}
\vspace{10mm}
\begin{abstract}
In the context of Calabi-Yau string models 
we explore the origin of characteristic pattern 
of quark-lepton masses and the CKM matrix. 
The discrete $R$-symmetry $Z_K \times Z_2$ 
is introduced and 
the $Z_2$ is assigned to the $R$-parity. 
The gauge symmetry at the string scale, 
$SU(6) \times SU(2)_R$, 
is broken into the standard model gauge group 
at a very large intermediate energy scale. 
At energies below the intermediate scale 
down-type quarks and also leptons are mixed 
with unobserved heavy states, respectively. 
On the other hand, there are no such mixings 
for up-type quarks. 
Due to the large mixings between light states 
and heavy ones we can derive phenomenologically 
viable fermion mass hierarchies and the CKM matrix. 
Mass spectra for intermediate-scale matter 
beyond the MSSM are also determined. 
Within this framework proton lifetime is long 
enough to be consistent with experimental data. 
As for the string scale unification of 
gauge couplings, however, 
consistent solutions are not yet found. 
\end{abstract} 

\newpage 


\section{Introduction} 
\hspace*{\parindent} 
In order to make sure of the reality of 
string theory it is important to explore 
how string theory determines low-energy 
parameters which are free parameters 
in the standard model and 
in the minimal supersymmetric standard model 
(MSSM). 
Among many issues of low-energy parameters, 
the characteristic pattern of quark-lepton 
masses and mixing angles has long been 
a challenging problem to explain its origin. 
The observed masses of quarks and leptons have 
the hierarchical pattern 
\begin{description} 
\item [(i)] \qquad \qquad $m$(1\,st gen.) 
          $\ll $ $m$(2\,nd gen.) $\ll $ $m$(3\,rd gen.) 
\end{description} 
and also the ratios among quark masses 
are in line as 
\begin{description} 
\item [(ii)] \qquad \qquad $m_u/m_d 
                 < m_c/m_s < m_t/m_b$. 
\end{description} 
Up to now several possibilities of explaining 
these features have been studied by many authors
\cite{Frog}-\cite{Infra}. 
A possibility is that all the observed pattern 
of fermion masses are attributable to 
the boundary condition, 
i.e. to the hierarchical structure of Yukawa 
couplings themselves at a very large scale. 
However, when we take GUT-type models, 
it is difficult to find a satisfactory solution 
in which property (ii) comes into line with 
a simple unification of Yukawa couplings. 
In this paper we explore a somewhat distinct 
possibility. 
In the context of Calabi-Yau 
string models with Kac-Moody level-one 
we propose a new type of model 
which potentially generates the characteristic 
pattern of fermion masses and 
Cabibbo-Kobayashi-Maskawa(CKM) matrix. 
In the model property (i) is attributed to 
the texture of renormalizable and nonrenormalizable 
interactions restricted by some discrete symmetries 
at the string scale. 
This mechanism is similar to those 
proposed in Refs.
\cite{Frog}\cite{Bound}\cite{LBL}\cite{Orbi}. 
On the other hand, property (ii) comes from 
large mixings among states observed at low energies 
and unobserved heavy ones. 
The mixings occur only for down-type quarks and 
for leptons below the energy scale at which 
the gauge group is broken into 
the standard model gauge group 
$G_{st} = SU(3)_c \times SU(2)_L \times U(1)_Y$.

The four-dimensional effective theory from 
Calabi-Yau string compactification 
is far more constrained than ordinary field theory. 
In the effective theory 
there are many peculiar features beyond the MSSM. 
First point is that the gauge group $G$, 
which is given via the flux breaking at 
the string scale $M_S$, 
is a subgroup of $E_6$ and 
would be larger than the standard model 
gauge group $G_{st}$. 
We will choose 
$G = SU(6) \times SU(2)_R$, 
under which doublet Higgs and color-triplet 
Higgs fields transform differently
\cite{Aligned}. 
As we will see later, the gauge group $G$ is 
spontaneously broken to $G_{st}$ in two steps 
at very large intermediate energy scales. 
Second point is that 
the massless sector of the Calabi-Yau string model 
contains extra particles beyond the MSSM. 
In string inspired models we typically have 
a number of generations and anti-generations. 
For illustration, if the gauge group $G$ is $E_6$, 
the massless chiral superfields apart from 
$E_6$-singlets consist of 
\begin{equation} 
       N_f \,{\bf 27} \ 
               + \ \delta \,({\bf 27} + {\bf 27^*})\,, 
\end{equation} 
where $N_f$ means the family number at low energies. 
It should be noted that $\delta $ sets of 
vector-like multiplets are included in 
the massless sector. 
In Calabi-Yau string compactification the generation 
structure of matter fields is 
closely linked to the topological structure 
of the compactified manifold. 
We will assume $\delta = 1$ 
for the sake of simplicity. 
Adjoint Higgs representations 
which are introduced in the traditional GUT 
are not allowed at Kac-Moody level-one. 
In addition, particles beyond the MSSM are 
contained also in ${\bf 27}$-representation of $E_6$. 
Namely in ${\bf 27}$ we have 
quark superfields $Q = (U, D)$, $U^c$, $D^c$, 
lepton superfields $L = (N, E)$, $N^c$, $E^c$, 
Higgs doublets $H_u$, $H_d$, 
color-triplet Higgses $g$, $g^c$ and 
an $SO(10)$-singlet $S$. 
When the gauge group $G$ is broken into $G_{st}$, 
superfields $D^c$ and $g^c$ as well as 
$L$ and $H_d$ are indistinguishable from 
each other under $G_{st}$.
Hence there possibly appear mixings between $D^c$ 
and $g^c$ and between $L$ and $H_d$. 
On the other hand, for up-type quarks there appear 
no such mixings. 
While up-type, down-type quarks and leptons 
share their interactions in common at 
the string scale $M_S$, 
$D^c$-$g^c$ mixing and $L$-$H_d$ mixing potentially 
turn mass pattern of down-type quarks 
and leptons out of that of up-type quarks 
at low energies. 
Further the mixings may be responsible 
for the CKM matrix. 
Third point of peculiar features beyond the MSSM is 
that superstring theory naturally provides 
the discrete symmetries 
which stem from symmetric structure of the 
compactified space. 
As shown in Gepner model 
\cite{Gepner}, 
the discrete symmetry can be the $R$-symmetry 
under which the components of a given superfield 
transform differently. 
Also, the discrete symmetry could be used 
as a horizontal symmetry. 
The discrete $R$-symmetry strongly limits 
the renormalizable and nonrenormalizable interactions 
and then possibly controls parameters in the 
low-energy effective theory. 
Recently it has been argued that the discrete 
$R$-symmetry controls energy scales of 
the symmetry breaking
\cite{Disc}, 
the magnitude of Majorana masses of 
the right-handed neutrino
\cite{Majorana} 
and the stability of the weak-scale hierarchy
\cite{Tadpole}. 
We will introduce the discrete $R$-symmetry 
$Z_K \times Z_2$ at the string scale.

In this paper main emphasis is placed on 
how both of the mixing mechanism mentioned above 
and the discrete symmetry bring about 
phenomenologically viable fermion mass pattern 
and the CKM matrix. 
This paper is organized as follows. 
In section 2 we introduce the discrete $R$-symmetry, 
which puts stringent constraints on 
interactions in the superpotential. 
The $Z_2$ symmetry is chosen so as to be in 
accord with the so-called $R$-parity in the MSSM. 
$R$-parity is conserved 
over the whole energy range from the string scale 
to the electroweak scale. 
After arguing that the discrete $R$-symmetry 
controls energy scales of the gauge 
symmetry breaking, 
we study particle spectra of vector-like 
multiplets in section 3. 
Since doublet Higgses and color-triplet Higgses 
belong to the different representations of 
the gauge group $G = SU(6) \times SU(2)_R$, 
distinct particle spectra of these fields 
are derived without some fine-tuning of 
parameters. 
In section 4 mass matrices for colored chiral 
multiplets are presented. 
There appear mixings between $D^c$ and $g^c$. 
Choosing appropriate assignments of discrete charges, 
we get large mixings between them. 
Due to the maximal mixing the mass pattern of 
down-type quarks differs from 
that of up-type quarks. 
The model generates not only 
the hierarchical pattern of quark masses 
but also the texture of the CKM matrix. 
In section 5 we discuss mixings between $L$ and 
$H_d$ and study spectra of leptons. 
The CKM matrix for leptons turns out to be 
a unit matrix. 
In the present framework we have several 
$R$-parity even colored superfields 
which potentially mediate proton decay. 
In section 6 it is shown that proton lifetime 
is about $10^{33 \sim 35}$year. 
In section 7 we find that the gauge coupling 
unification is not successfully achieved 
as a consequence of spectra of 
extra intermediate-scale matter. 
In the final section we conclude with a brief 
summary of our results. 
In Appendix A it is shown that 
under an appropriate condition on 
the soft SUSY breaking parameters 
the gauge symmetry is broken at tree level. 
In Appendix B we show that if neutrino Majorana 
masses are sufficiently large compared with 
the soft supersymmetry(SUSY) breaking scale 
$m_{3/2} = O(1{\rm TeV})$, 
the scalar potential is minimized along the 
direction where $R$-parity is conserved.


\setcounter{equation}{0}
\section{Discrete $R$-symmetry} 
\hspace*{\parindent} 
In order to guarantee the stability of 
the weak-scale hierarchy without fine-tuning, 
it is favorable that doublet Higgses and color-triplet 
Higgses reside in different irreducible 
representations of the string scale gauge group $G$. 
As the largest gauge group implementing such 
a situation is $G = SU(6) \times SU(2)_R$
\cite{Aligned}, 
in this paper we choose $SU(6) \times SU(2)_R$ 
as an example of $G$. 
Chiral superfields $(\Phi )$ in ${\bf 27}$ 
representation of $E_6$ are decomposed into 
\begin{eqnarray*} 
\Phi ({\bf 15, 1})  &:& \ \ Q, L, g, g^c, S, \\
\Phi ({\bf 6^*, 2}) &:& \ \ U^c, D^c, N^c, E^c, 
                                      H_u, H_d. 
\end{eqnarray*} 
Although $L$ and $H_d$ ($D^c$ and $g^c$) 
have the same quantum numbers under $G_{st}$, 
they belong to different irreducible 
representations of $SU(6) \times SU(2)_R$. 
The superpotential $W$ is described in terms of 
${\bf 27}$ chiral superfields $(\Phi )$ and 
${\bf 27^*}$ ones $(\overline {\Phi })$ as 
\begin{equation} 
     W = \Phi ^3 + \overline {\Phi}^3 + 
            (\Phi \overline {\Phi })^{m+1} + 
                \Phi ^3 (\Phi \overline {\Phi })^n 
                   + \cdots , 
\end{equation} 
where $m$ and $n$ are positive integers and 
all the terms are characterized by 
the couplings of $O(1)$ in units of 
$M_S = O(10^{18}{\rm GeV})$. 
The cubic term $\Phi ^3$ is of the forms 
\begin{eqnarray} 
    (\Phi ({\bf 15, 1}))^3 & = & QQg + Qg^cL + g^cgS, \\
    \Phi ({\bf 15, 1})(\Phi ({\bf 6^*, 2}))^2 & 
            = & QH_dD^c + QH_uU^c + LH_dE^c  + LH_uN^c 
                                            \nonumber \\ 
             {}& & \qquad   + SH_uH_d + 
                     gN^cD^c + gE^cU^c + g^cU^cD^c.
\end{eqnarray} 

We assume that the massless matter fields are 
composed of chiral multiplets 
$\Phi _i$ $(i = 1,\cdots , N_f=3)$ and a set 
$(\delta = 1)$ of vector-like multiplets $\Phi _0$ 
and $\overline {\Phi }$. 
Here we introduce the discrete $R$-symmetry 
$Z_K \times Z_2$ as a stringy selection rule. 
As we will see below, large $K$ is favorable 
for explaining the mass pattern of quarks 
and leptons. 
The $Z_2$ symmetry is taken so as to be in accord 
with the $R$-parity in the MSSM. 
Therefore, hereafter the $Z_2$ symmetry is referred 
to as $R$-parity. 
Supposing that ordinary quarks and leptons are 
included in chiral multiplets 
$\Phi _i$ $(i = 1,2,3)$, 
$R$-parity of all 
$\Phi _i$ $(i = 1,2,3)$ are set to be odd. 
Since light Higgs scalars are even under $R$-parity, 
light Higgs doublets are bound to reside in 
$\Phi _0$ and/or $\overline {\Phi }$. 
For this reason we assign even $R$-parity 
to $\Phi _0$ and $\overline {\Phi }$. 
In Appendix B we show that 
once the $R$-parity is conserved 
at the string scale, 
the $R$-parity remains unbroken down to 
the electroweak scale under appropriate conditions. 
Hence, through the spontaneous breaking of gauge 
symmetry gauge superfields are possibly mixed with 
the vector-like multiplets $\Phi _0$ and 
$\overline {\Phi }$
but not with the chiral multiplets $\Phi _i$ 
$(i=1,2,3)$. 
Furthermore, no mixing occurs between the vector-like 
multiplets and the chiral multiplets.

We use the $Z_K$ symmetry as a horizontal 
symmetry and construct our model incorporating 
the mechanism of Froggatt and Nielsen\cite{Frog}. 
The $Z_K$ symmetry controls 
not only a large hierarchy of 
the energy scales of the symmetry breaking 
but also the texture of effective Yukawa couplings. 
We denote the $Z_K$-charges of chiral multiplets 
$\Phi _i({\bf 15, 1})$ and $\Phi _i({\bf 6^*, 2})$ 
by $a_i$ and $b_i$ $(i = 0, 1, 2, 3)$, 
respectively. 
In Table I, we tabulate the notations for 
$Z_K$-charges and the assignment of $R$-parity 
for each superfield. 
Note that the anticommuting number $\theta $ has 
also a $Z_K \times Z_2$-charge $(-1, -)$.

\vspace {5mm} 
\begin{center} 
\framebox [3cm] {\large \bf Table I} 
\end{center} 
\vspace {5mm} 


\setcounter{equation}{0}
\section{Gauge hierarchy and the $\mu $-term} 
\hspace*{\parindent} 
The discrete symmetry introduced above puts 
stringent constraints on both renormalizable 
and nonrenormalizable interactions 
in the superpotential. 
To begin with, $Z_K$-charges of 
vector-like multiplets are chosen such that 
both the nonrenormalizable interactions 
\begin{equation} 
   \left( \Phi _0{\bf (15, 1)} 
       \overline {\Phi }{\bf (15^*, 1)} \right)^{sk} 
\end{equation} 
and 
\begin{equation} 
   \left( \Phi _0{\bf (6^*, 2)} 
       \overline {\Phi }{\bf (6, 2)} \right)^s 
\end{equation} 
are allowed, 
where $K=sk+1$ and $s$ and $k$ are even and odd 
integers larger than unity, respectively. 
This implies that 
$sk(a_0 + {\overline a}) + 2 \equiv 
s(b_0 + {\overline b}) + 2 \equiv 0$ in modulus 
$K=sk+1$. 
Thus we impose 
\begin{equation} 
   a_0 + {\overline a} \equiv 2, \qquad 
   b_0 + {\overline b} \equiv 2k  \qquad 
                                {\rm mod}\ K. 
\label{eqn:ab} 
\end{equation} 
It follows that the interactions 
\begin{equation} 
  W_{SN} \sim \sum_{r=0}^{s}
       (\Phi _0{\bf (15, 1)} 
        {\overline \Phi}{\bf (15^*, 1)})^{(s-r)k} 
            (\Phi _0{\bf (6^*, 2)} 
             {\overline \Phi}{\bf (6, 2)})^r 
\end{equation} 
are allowed in $M_S$ units. 
Due to $R$-parity conservation 
the interactions containing even number of 
$\Phi _i$ $(i=1, 2, 3)$ are also allowed but 
all of the interactions containing odd number 
of $\Phi _i$ $(i=1, 2, 3)$ are forbidden.

Incorporating the soft SUSY breaking 
terms together with the $F$- and $D$-terms, 
we get the scalar potential. 
Although dynamics of SUSY breaking is not 
presently known, 
we may parametrize the SUSY breaking by 
introducing the universal soft terms. 
The scale of SUSY breaking $m_{3/2}$ is 
supposed to be $O(1{\rm TeV})$. 
Through the minimization of the scalar potential 
we are able to detemine a ground state, 
which is characterized by VEVs of 
$\Phi _0$, $\overline \Phi $ and 
$\Phi _i$ $(i=1, 2, 3)$. 
Under appropriate conditions on soft SUSY breaking 
parameters the gauge symmetry is spontaneously 
broken at tree level (see Appendix A). 
Further, if masses of $G_{st}$-neutral and 
$R$-parity odd superfields are sufficiently larger 
than $m_{3/2}$, 
the scalar potential is minimized at vanishing 
$\langle \Phi _i \rangle$ for $i=1, 2, 3$ 
(see Appendix B). 
On the other hand, 
$\Phi _0$ and ${\overline \Phi }$ 
acquire nonzero VEVs along a $D$-flat 
direction, namely 
\begin{eqnarray} 
     \langle \Phi _0{\bf (15, 1)} \rangle & = & 
     \langle {\overline \Phi}{\bf (15^*, 1)} \rangle \ 
                 \simeq \  M_S \,x, \\
     \langle \Phi _0{\bf (6^*, 2)} \rangle & = & 
     \langle {\overline \Phi}{\bf (6, 2)} \rangle \ 
                 \simeq \  M_S \,x^k 
\end{eqnarray} 
up to phase factors
\cite{Disc}\cite{Majorana}, 
where 
\begin{equation} 
     x = \left ( \frac{m_{3/2}}{M_S} \right )
                             ^{1/(2sk-2)}. 
\label{eqn:x} 
\end{equation} 
Although for a large $K$ the parameter $x$ 
by itself is not a very small number, 
the large hierarchy occurs by raising the number 
to large powers. 
Hence, $x$ becomes an efficient parameter 
in describing the hierarchical structure 
of the effective theory. 
Note that we have the inequalities 
\begin{equation} 
   M_S > |\langle \Phi _0({\bf 15, 1}) \rangle | > 
      |\langle \Phi _0({\bf 6^*, 2}) \rangle | \gg 
                             \sqrt{m_{3/2}M_S}. 
\end{equation} 
Hereafter the fields $\Phi _0{\bf (15, 1)}$ and 
${\overline \Phi }{\bf (15^*, 1)}$ which 
develop non-zero VEVs are referred to as 
$G_{st}$-neutral fields $S_0$ and ${\overline S}$, 
respectively. 
At the scale $\langle S_0 \rangle = 
\langle {\overline S} \rangle \simeq M_S\,x $ 
the gauge symmetry $SU(6) \times SU(2)_R$ 
is spontaneously broken to 
$SU(4)_{PS} \times SU(2)_L \times SU(2)_R$, 
where $SU(4)_{PS}$ stands for the Pati-Salam $SU(4)$
\cite{Pati}. 
Under the $SU(4)_{PS} \times SU(2)_L \times SU(2)_R$ 
the chiral superfields $\Phi {\bf (15, 1)}$ and 
$\Phi {\bf (6^*, 2)}$ are decomposed as 
\begin{eqnarray}
   {\bf (15, 1)} &=& {\bf (4,2,1)} + {\bf (6,1,1)} 
                      + {\bf (1,1,1)}, \\
   {\bf (6^*, 2)} &=& {\bf (4^*,1,2)} + {\bf (1,2,2)}, 
\end{eqnarray}
where each matter field is assigned as 
\begin{eqnarray*} 
\Phi ({\bf 4,2,1})   &:& \ \ Q, L, \\
\Phi ({\bf 6,1,1})   &:& \ \ g, g^c, \\
\Phi ({\bf 1,1,1})   &:& \ \ S, \\
\Phi ({\bf 4^*,1,2}) &:& \ \ U^c, D^c, N^c, E^c, \\
\Phi ({\bf 1,2,2})   &:& \ \ H_u, H_d. 
\end{eqnarray*} 

The subsequent symmetry breaking takes place 
via the non-zero VEVs 
$\langle \Phi _0{\bf (6^*, 2)} \rangle $ $=$ 
$\langle {\overline \Phi }{\bf (6, 2)} \rangle $ 
$\simeq M_S\,x^k $. 
At this stage of the symmetry breaking there seem 
to be two possibilities depending on whether 
the non-zero VEV 
$\langle \Phi _0{\bf (6^*, 2)} \rangle 
(\langle {\overline \Phi }{\bf (6, 2)} \rangle )$ 
is attributed to 
$\langle \Phi _0{\bf (4^*,1,2)} \rangle 
(\langle {\overline \Phi }{\bf (4,1,2)} \rangle )$ 
or 
$\langle \Phi _0{\bf (1,2,2)} \rangle 
(\langle {\overline \Phi }{\bf (1,2,2)} \rangle )$. 
As will be seen soon later, 
we have the term $(S_0 {\overline S})^p{\overline S}
{\overline H_u}{\overline H_d}$ in the superpotential, 
where $p$ is a positive integer detemined by 
the discrete symmetry $Z_K$. 
Under an appropriate charge assignment of 
matter fields we have $p \simeq sk-2k$. 
According to the presence of this superpotential 
term the large VEV 
$\langle {\overline \Phi}{\bf (1,2,2)} \rangle $
is inconsistent with the (almost) $F$-flat condition. 
Consequently, the subsequent symmetry breaking 
occurs through 
$\langle \Phi _0{\bf (4^*,1,2)} \rangle 
= \langle {\overline \Phi }{\bf (4,1,2)} \rangle 
\simeq M_S\,x^k $. 
Then we denote the fields $\Phi _0{\bf (4^*,1,2)}$ 
and ${\overline \Phi }{\bf (4,1,2)}$ 
with the non-zero VEVs as $N^c_0$ and 
${\overline N^c}$, respectively. 
Thus the gauge symmetry is spontaneously broken 
in two steps at the scales $\langle S_0 \rangle $ 
and $\langle N^c_0 \rangle$ as 
\begin{eqnarray} 
   SU(6) \times SU(2)_R 
   & \buildrel \langle S_0 \rangle \over \longrightarrow &
             SU(4)_{PS} \times SU(2)_L \times SU(2)_R \\
   & \buildrel \langle N^c_0 \rangle \over \longrightarrow &
             SU(3)_c \times SU(2)_L \times U(1)_Y. 
\end{eqnarray} 
Since $S_0$, ${\overline S}$, $N^c_0$ and 
${\overline N^c}$ acquire VEVs along a $D$-flat 
direction, 
SUSY is maintained down to $O(1{\rm TeV})$.

At the first step of the symmetry breaking 
chiral superfields $Q_0$, $L_0$, ${\overline Q}$, 
${\overline L}$ and $(S_0 - {\overline S})/\sqrt{2}$ 
are absorbed by gauge superfields. 
Through the subsequent symmetry breaking 
chiral superfields $U^c_0$, $E^c_0$, ${\overline U^c}$, 
${\overline E^c}$ and $(N^c_0 - {\overline N^c})/\sqrt{2}$ 
are absorbed. 
On the other hand, for components $(S_0 + {\overline S})/\sqrt{2}$ 
and $(N^c_0 + {\overline N^c})/\sqrt{2}$ 
the mass matrix is of the form 
\begin{equation} 
   \left(
   \begin{array}{cc}
      O(x^{2sk-2})   &  O(x^{(2s-1)k-1}) \\
      O(x^{(2s-1)k-1})   &  O(x^{2(s-1)k}) 
   \end{array}
   \right) 
\end{equation} 
in $M_S$ units. 
This yields mass eigenvalues 
\begin{equation} 
   O(m_{3/2}), \ \ O(M_S x^{2(s-1)k}), 
\end{equation} 
which correspond to the eigenstates 
\begin{equation} 
\begin{array}{c} 
   \vphantom{\bigg(} 
   \frac {1}{\sqrt{2}}(S_0 + {\overline S}) 
       + O(x^{k-1})\frac {1}{\sqrt{2}}
                  (N^c_0 + {\overline N^c}),  \\ 
   \vphantom{\bigg(} 
   \frac {1}{\sqrt{2}}(N^c_0 + {\overline N^c}) 
       + O(x^{k-1})\frac {1}{\sqrt{2}}
                  (S_0 + {\overline S}), 
\end{array}
\end{equation} 
respectively
\cite{Majorana}. 
The discrete symmetry $Z_K$ is 
broken together with $SU(6) \times SU(2)_R$ 
by the VEV $\langle S_0 \rangle $, 
while the VEV allows the $Z_2$-symmetry 
(referred to $R$-parity conservation) 
to remain unbroken all the way down to TeV.

In order to stabilize the weak-scale hierarchy 
we put an additional requirement that 
the interaction 
\begin{equation} 
   (S_0 {\overline S})^{sk-e} S_0 H_{u0} H_{d0} 
\end{equation} 
is allowed with $e = 0,\, 1$ in the superpotential. 
We will shortly show that the $\mu $-problem is 
solved by this setting $e = 0,\, 1$. 
This condition is translated into 
\begin{equation} 
     a_0 + 2b_0 \equiv  2e \ \ \ {\rm mod}\ K. 
\label{eqn:abb} 
\end{equation} 
From Eqs.(\ref{eqn:ab}) and (\ref{eqn:abb}) 
the superpotential of Higgs doublet 
in vector-like multiplets has the form 
\begin{eqnarray} 
    W_H  & \sim &  (S_0 {\overline S})^{(s-2)k+e-1} 
         {\overline S}\,{\overline H_u}{\overline H_d} 
              + (S_0 {\overline S})^{(s-1)k} 
                           (H_{u0} {\overline H_u} 
                 + H_{d0} {\overline H_d}) \nonumber  \\ 
         &  & \hphantom{QQQ} + (S_0 {\overline S})^{sk-e} 
                              S_0 H_{u0} H_{d0}. 
\end{eqnarray} 
When $S_0$ and $\overline S$ develop the non-zero 
VEVs, 
the superpotential induces the mass matrix 
of $H_{u0}$, $H_{d0}$, 
${\overline H_u}$ and ${\overline H_d}$ 
\begin{equation} 
\begin{array}{r@{}l} 
    \vphantom{\bigg(}  &  \begin{array}{ccc} 
        \qquad \qquad {\overline H_u} \qquad & \qquad H_{d0} & 
       \end{array} \\ 
    \begin{array}{l} 
      {\overline H_d}  \\  H_{u0} \\
    \end{array} 
    & 
 \left( 
   \begin{array}{cc}
      O(x^{2(s-2)k+2e-1})   &  O(x^{2(s-1)k})   \\
      O(x^{2(s-1)k})        &  O(x^{2sk-2e+1}) 
   \end{array}
 \right) 
\end{array} 
\end{equation} 
in $M_S$ units, 
which leads to the mass eigenvalues 
\begin{equation} 
   O(M_S \,x^{2(s-2)k+2e-1}), \ \ 
       O(M_S \,x^{2sk-2e+1}) = O(m_{3/2} \,x^{3-2e}).
\end{equation} 
Consequently, we have the $\mu $-term with $m_{3/2} > \mu = 
O(m_{3/2} \,x^{3-2e}) = O(10^{2 \sim 3}{\rm GeV})$ 
for $e=0,1$
\cite{Aligned}\cite{Tadpole}. 
Here, note that we take $x \sim 0.7$ with 
$sk = 50$ in a typical example given later. 
Light Higgs states are given by 
\begin{equation} 
     H_{u0} + O(x^{2k+1-2e}){\overline H_d}, \qquad 
     H_{d0} + O(x^{2k+1-2e}){\overline H_u}. 
\end{equation} 
The components of $\overline{H_d}$ and 
$\overline{H_u}$ in light Higgses are small. 
Generally speaking, in the superpotential $W_H$ 
there exist additional terms which are obtained 
by replacing each factor $(S_0{\overline S})^k$ 
by a factor $(N^c_0{\overline N^c})$. 
However, as far as the mass matrices are concerned, 
these terms yield the same order of magnitude as 
in each entry of the above matrix because of the 
relations $(|\langle S_0 \rangle |/M_S)^k = 
|\langle N^c_0 \rangle |/M_S$ and 
$k(a_0+{\overline a}) \equiv b_0+{\overline b}$. 
Since we do not address here the issue of 
CP-violation, 
all VEVs are assumed to be real for simplicity. 
Therefore, hereafter the nonrenormalizable 
terms are expressed in terms only of the 
powers of $(S_0{\overline S})$. 
Note that the product $H_{u0} H_{d0}$ has a nonzero 
$Z_K$-charge. 
In contrast with the present model, 
in a solution of the $\mu $-problem proposed in 
Ref.\cite{Miu} 
the $R$-charge of the product of light Higgses 
has to be zero.

The remaining components in $\Phi _0$ and 
$\overline {\Phi }$, i.e. $g_0$, $g^c_0$, $D^c_0$ and 
$\overline g$, ${\overline g^c}$, ${\overline D^c}$ 
are down-type color-triplet fields. 
In the present model the spectra of color-triplet 
Higgses are quite different from those of 
doublet Higgses. 
Mass matrix for these fields is given 
in section 6.


\setcounter{equation}{0}
\section{Quark masses and the CKM matrix} 
\hspace*{\parindent} 
Next we turn to mass matrices for chiral multiplets 
$\Phi _i$ $(i=1, 2, 3)$. 
Due to $R$-parity conservation 
$\Phi _i$ $(i= 1, 2, 3)$ 
are not mixed with vector-like multiplets 
$\Phi _0$ and $\overline {\Phi }$. 
The superpotential of up-type quarks 
which contributes to the mass matrix 
of up-type quarks, is given by 
\begin{equation} 
   W_U \sim (S_0 \overline {S})^{m_{ij}} Q_i U^c_j H_{u0} 
          \qquad (i, j = 1, 2, 3), 
\end{equation} 
where the exponents $m_{ij}$ are integers 
in the range $0 \leq m_{ij} < K=sk+1$. 
Although the $Z_K$ symmetry allows the terms 
multiplied by $(S_0 {\overline S})^K$, 
$(S_0 {\overline S})^{2K}$, $\cdots $, 
the contributions of these terms are negligibly 
small compared with the above ones. 
Therefore, it is sufficient for us to 
take only the terms with $m_{ij} < K$. 
Recall that light Higgs doublets are almost 
$H_{u0}$ and $H_{d0}$. 
Under the $Z_K$-symmetry the exponent $m_{ij}$ 
is determined by the condition 
\begin{equation} 
   2 m_{ij} + a_i + b_j + b_0 + 2 \equiv 0 
                        \qquad {\rm mod}\ K. 
\label{eqn:mab}
\end{equation} 
Instead of $a_i$ and $b_i$ $(i = 1, 2, 3)$, 
hereafter we introduce new notations 
$\alpha ,\ \beta ,\ \gamma ,$ and $\delta $ 
defined by 
\begin{equation} 
    a_2 - a_1  \equiv \alpha , \quad 
    b_2 - b_1  \equiv \beta ,  \quad
    a_3 - a_2  \equiv \gamma , \quad 
    b_3 - b_2  \equiv \delta . 
\end{equation} 
These parameters are supposed to be even integers 
to derive desirable mass pattern of quarks 
and 
\begin{equation}
    0 < \alpha \leq \delta \leq \gamma \leq \beta ,  
                    \qquad    2(\beta + \delta )< K. 
\end{equation}
The above condition (\ref{eqn:mab}) 
is rewritten as 
\begin{equation} 
   2 m_{ij} \equiv 2m_{33} + \left(
      \begin{array}{ccc}
        \alpha + \beta + \gamma + \delta  &  
                \alpha + \gamma + \delta  &  
                      \alpha + \gamma   \\ 
        \beta + \gamma + \delta  &  
              \gamma + \delta  &  \gamma  \\ 
        \beta + \delta  &  \delta  &  0   \\
      \end{array} 
             \right)_{ij}      \qquad {\rm mod}\ K, 
\label{eqn:mij} 
\end{equation} 
where $2m_{33} \equiv -a_3-b_3-b_0-2$. 
The mass matrix of up-type 
quarks is described by a $3 \times 3$ matrix 
$M$ with elements 
\begin{equation} 
         M_{ij} = O(x^{2 m_{ij}}) 
\end{equation} 
multiplied by $v_u = \langle H_{u0} \rangle $. 
This equation is an order of magnitude relationship, 
so that each element will be multiplied by an 
$O(1)$ number. 
From Eq.(\ref{eqn:mij}) 
the matrix $M$ is generally asymmetric. 
Here we take an ansatz that only top-quark has 
a trilinear coupling. 
This means that 
\begin{equation}
     m_{33} = 0. 
\end{equation}
When we adopt appropriate unitary matrices 
${\cal V}_u$ and ${\cal U}_u$, 
the matrix 
\begin{equation} 
    {\cal V}_u^{-1} M \,{\cal U}_u 
\end{equation} 
becomes diagonal. 
Explicitly, ${\cal V}_u$ and ${\cal U}_u$ 
are of the forms 
\begin{eqnarray} 
    {\cal V}_u & = & 
      \left( 
      \begin{array}{ccc}
        1 - O(x^{2\alpha })  &  O(x^{\alpha})  
                          &  O(x^{\alpha + \gamma })  \\
        O(x^{\alpha })    &  1 - O(x^{2\alpha })  
                                   &  O(x^{\gamma })  \\
        O(x^{\alpha + \gamma })   &  O(x^{\gamma })  
                              &  1 - O(x^{2\gamma })  \\
      \end{array}
      \right), \\
    {\cal U}_u & = & 
      \left( 
      \begin{array}{ccc}
        1 - O(x^{2\beta })  &  O(x^{\beta})  
                          &  O(x^{\beta + \delta })  \\
        O(x^{\beta })    &  1 - O(x^{2\delta })  
                                   &  O(x^{\delta })  \\
        O(x^{\beta + \delta })   &  O(x^{\delta })  
                              &  1 - O(x^{2\delta })  \\
      \end{array}
      \right). 
\end{eqnarray} 
The eigenvalues of $M$ are 
\begin{equation}
   O(x^{\alpha + \beta + \gamma + \delta }), \qquad 
   O(x^{\gamma + \delta }), \qquad  O(1), 
\end{equation}
which correspond to $u$-, $c$- and $t$-quarks,
respectively.

Under $SU(6) \times SU(2)_R$ gauge symmetry 
down-type quarks and leptons share 
the nonrenormalizable terms in common with 
up-type quarks. 
Namely we get 
\begin{equation} 
   W \sim (S_0 \overline {S})^{m_{ij}} 
       \{ Q_i D^c_j H_{d0} + L_i N^c_j H_{u0} 
           + L_i E^c_j H_{d0} \}. 
\end{equation} 
For down-type quarks, however, the mixings between 
$g^c$ and $D^c$ should be taken into account 
at energies below the scale $\langle N_0^c \rangle$. 
This is because we have two down-type 
$SU(2)_L$-singlet colored fields in each 
${\bf 27}$ of $E_6$. 
Then, hereafter we denote $R$-parity odd 
$g_i$ and $g^c_i$ $(i=1,2,3)$ as $D'_i$ and 
$D'^c_i$ $(i=1,2,3)$, respectively. 
The superpotential of down-type colored 
fields is of the form 
\begin{equation} 
   W_D \sim (S_0 \overline {S})^{z_{ij}} S_0 D'_i D'^c_j 
           + (S_0 \overline {S})^{m_{ij}} 
       ( N^c_0 D'_i + H_{d0} Q_i) D^c_j, 
\end{equation} 
where the exponents $z_{ij}$ are determined by 
\begin{equation} 
    2 z_{ij} + a_i + a_j + a_0 +2 \equiv 0  \qquad 
                           {\rm mod}\ K 
\label{eqn:zaa}
\end{equation} 
in the range $0 \leq z_{ij} < K$. 
Thus we have 
\begin{equation} 
   2 z_{ij} \equiv 2z_{33} + \left(
      \begin{array}{ccc}
        2\alpha + 2\gamma  &  \alpha + 2\gamma  &  
                              \alpha + \gamma    \\ 
        \alpha + 2\gamma  &  2\gamma  &  \gamma  \\ 
        \alpha + \gamma   &   \gamma  &     0    \\
      \end{array} 
             \right)_{ij}      \qquad {\rm mod}\ K 
\label{eqn:zij} 
\end{equation} 
with $2z_{33} \equiv -2a_3 - a_0 - 2$. 
In terms of a $3 \times 3$ matrix $Z$ 
with elements 
\begin{equation} 
     Z_{ij} = O(x^{2z_{ij}}), 
\end{equation} 
a mass matrix of down-type colored fields is 
written as 
\begin{equation} 
\begin{array}{r@{}l} 
   \vphantom{\bigg(}   &  \begin{array}{ccc} 
          \quad   D'^c   &  \quad D^c  &  
        \end{array}  \\ 
\widehat{M}_d = 
   \begin{array}{l} 
        D'  \\  D  \\ 
   \end{array} 
     & 
\left( 
  \begin{array}{cc} 
      x Z   &    x^k M    \\
       0    &  \rho _d M 
  \end{array} 
\right) 
\end{array} 
\label{eqn:Mdh} 
\end{equation} 
in $M_S$ units below the scale 
$\langle N^c_0 \rangle $, 
where $\rho _d = \langle H_{d0} \rangle /M_S = v_d /M_S$. 
This $\widehat{M}_d$ is a $6 \times 6$ matrix 
and can be diagonalized by a bi-unitary transformation 
as 
\begin{equation} 
    \widehat{\cal V}_d^{-1} \widehat{M}_d \, 
                        \widehat{\cal U}_d. 
\label{eqn:Md} 
\end{equation} 
$\widehat{M}_d$ shows mixings between 
$D'^c$ and $D^c$, explicitly. 
This type of mixings does not occur for up-type 
quarks. 
From Eqs. (\ref{eqn:Mdh}) and (\ref{eqn:Md}) 
the matrix 
\begin{equation} 
    \widehat{\cal V}_d^{-1} \widehat{M}_d 
         \widehat{M}_d^{\dag } \widehat{\cal V}_d = 
    \widehat{\cal V}_d^{-1} 
      \left( 
      \begin{array}{cc}
          A_d + B_d    &  \epsilon _d B_d   \\
        \epsilon _d B_d   &  \epsilon _d^2 B_d 
      \end{array}
      \right)
    \widehat{\cal V}_d 
\end{equation} 
is diagonal, where 
\begin{equation} 
  A_d = x^2 Z Z^{\dag}, \qquad B_d = x^{2k} M M^{\dag}, 
                    \qquad \epsilon _d = \rho _d \,x^{-k}. 
\end{equation} 

In view of the smallness of the parameter 
$\epsilon _d $, 
we use the perturbative method 
in solving the eigenvalue problem. 
It follows that the eigen equation is approximately 
separated into two pieces. 
For heavy states the eigen equation becomes 
\begin{equation} 
   \det \left( A_d + B_d - \frac {\eta }{M_S^2}
                \right) = 0. 
\end{equation} 
Solving this equation of a variable $\eta $, 
we obtain masses squared for three heavy states. 
The other three states are light and their masses 
are given by solving the eigen equation 
\begin{equation} 
   \det \left( x^{-2k}( A_d^{-1} + B_d^{-1} )^{-1}
         - \frac {\eta }{v_d^2} \right) = 0. 
\end{equation} 
This equation is derived in $\epsilon _d^2$ order of 
the perturbative expansion. 
The light states correspond to observed down-type 
quarks. 
If the mixing between $D'^c$ and $D^c$ is sizable, 
mass pattern of down-type quarks is possibly 
changed from that of up-type quarks. 
Thus in our model, property (ii) pointed out 
in section 1 for observed fermion masses 
is attributable to this mixing mechanism.

The $6 \times 6$ unitary matrices 
$\widehat{\cal{V}}_d$ and $\widehat{\cal{U}}_d$ 
are 
\begin{eqnarray} 
   \widehat{\cal V}_d & \simeq & \left( 
   \begin{array}{cc} 
      {\cal W}_d   &  -\epsilon _d (A_d + B_d)^{-1} 
                                    B_d {\cal V}_d \\
      \epsilon _d B_d (A_d + B_d)^{-1} {\cal W}_d    &  
                                          {\cal V}_d 
   \end{array} 
                       \right), \\ 
   \widehat{\cal U}_d & \simeq & \left( 
   \begin{array}{cc} 
      xZ^{\dag} {\cal W}_d \,(\Lambda _d^{(0)})^{-1/2} 
         &  -(xZ)^{-1} {\cal V}_d \,
                               (\Lambda _d^{(2)})^{1/2}  \\
      x^k M^{\dag} {\cal W}_d \,(\Lambda _d^{(0)})^{-1/2} 
         &  (x^kM)^{-1} {\cal V}_d \,
                                 (\Lambda _d^{(2)})^{1/2} 
   \end{array} 
   \right), 
\label{eqn:hatud}
\end{eqnarray} 
respectively. 
Here ${\cal W}_d$ and ${\cal V}_d$ are 
$3 \times 3$ unitary matrices 
which are determined such that the matrices 
\begin{equation} 
    {\cal W}_d^{-1}(A_d + B_d){\cal W}_d 
                          = \Lambda _d^{(0)}, \qquad 
    {\cal V}_d^{-1}(A_d^{-1} + B_d^{-1})^{-1}
              {\cal V}_d = \Lambda _d^{(2)} 
\end{equation} 
become diagonal. 
As a consequence we can expect to 
have a nontrivial CKM matrix 
\begin{equation} 
   V^{CKM} = {\cal V}_u^{-1} \,{\cal V}_d. 
\end{equation} 
Note that ${\cal V}_u$ is determined such that 
${\cal V}_u^{-1} B_d {\cal V}_u$ is diagonal. 
If the relation 
\begin{equation} 
    | \det (A_d+B_d)| \simeq | \det A_d | 
                             \gg | \det B_d | 
\end{equation} 
is satisfied, 
the mixing is small and we have 
\begin{equation} 
    ( A_d^{-1} + B_d^{-1} )^{-1} \simeq B_d. 
\end{equation} 
This implies that mass pattern of down-type 
quarks is the same as that of up-type quarks 
and that ${\cal V}_d \simeq {\cal V}_u$. 
In this case $V^{CKM}$ becomes 
almost a unit matrix.

To get a phenomenologically viable solution, 
large mixings between $D'^c$ and $D^c$ 
are preferable. 
Thus we impose the maximal mixing in which 
$(A_d^{-1})_{ij}$ and $(B_d^{-1})_{ij}$ 
are the same order. 
The maximal mixing is realized under 
the condition 
\begin{equation}
    2z_{33} = k - 1 - \alpha - \gamma 
                       + \beta + \delta . 
\end{equation}
Note that $k$ is an odd integer. 
Under the above condition on $z_{33}$
the eigenvalues of $A_d + B_d$ become 
\begin{equation}
    O(x^{2(k + \alpha + \gamma + \delta )}), \qquad 
    O(x^{2(k + \beta +\delta -\alpha )}), \qquad 
    O(x^{2k}). 
\end{equation}
It follows that extra down-type heavy quarks 
have their masses 
\begin{equation}
  M_S\,x^{k + \alpha + \gamma +\delta }, \qquad 
  M_S\,x^{k + \beta + \delta -\alpha }, \qquad 
  M_S\,x^k. 
\end{equation}
Main components of these eigenstates are 
$D_1$-$(O(1)D^c_1+O(1)D^c_2)$, $D_2$-$D'^c_3$ 
and $D_3$-$D'^c_2$, respectively. 
On the other hand, down-type light quarks 
have their masses 
\begin{equation}
  v_d\,x^{\alpha + \beta + \gamma + \delta }, \qquad 
  v_d\,x^{\beta + \gamma + \delta}, \qquad 
  v_d\,x^{-\alpha + \beta + \delta}, 
\end{equation}
which correspond to observed $d$-, $s$- and $b$-quarks. 
These eigenstates are approximately 
$D_1$-$(O(1)D^c_1+O(1)D^c_2)$, $D_2$-$D'^c_1$ 
and $D_3$-$D'^c_3$, respectively. 
It should be noted that we have very large 
$D^c_i$-$D'^c_i$ mixings. 
The unitary matrix ${\cal V}_d$ which diagonalizes 
$A_d^{-1} + B_d^{-1}$, is expressed as 
\begin{equation} 
   {\cal V}_d = \left(
      \begin{array}{ccc}
        1 - O(x^{2\alpha })  &  O(x^{\alpha })  &  
                              O(x^{\alpha + \gamma })   \\ 
        O(x^{\alpha })  &  1 - O(x^{2\alpha })  &  
                              O(x^{\gamma })            \\ 
        O(x^{\alpha + \gamma })  &   O(x^{\gamma }) & 
                              1 - O(x^{2\gamma })    \\
      \end{array} 
             \right).
\label{eqn:Vd} 
\end{equation} 
Corresponding elements of the matrices 
${\cal V}_u$ and ${\cal V}_d$ are in the same 
order of magnitudes but their coefficients 
of the leading term in off-diagonal elements 
are different with each other because of 
the maximal mixing. 
Consequently, the CKM matrix is given by 
\begin{equation} 
   V^{CKM} = {\cal V}_u^{-1} {\cal V}_d = \left(
      \begin{array}{ccc}
        1 - O(x^{2\alpha })  &  O(x^{\alpha })  &  
                              O(x^{\alpha + \gamma })   \\ 
        O(x^{\alpha })  &  1 - O(x^{2\alpha })  &  
                              O(x^{\gamma })            \\ 
        O(x^{\alpha + \gamma })  &   O(x^{\gamma }) & 
                              1 - O(x^{2\gamma })    \\
      \end{array} 
             \right).
\label{eqn:CKM} 
\end{equation} 
It is worth noting that large $D^c_i$-$D'^c_i$ 
mixings play an essential role in generating 
a nontrivial CKM matrix. 
An early attempt of explaining the CKM matrix 
via $D^c_i$-$D'^c_i$ mixings has been made 
in Ref.\cite{Hisano}, 
in which a SUSY $SO(10)$ model was taken.

Confronting the CKM matrix obtained here 
with the observed one, 
it is feasible for us to take 
\begin{equation}
    \alpha = 1.0 \times w, \qquad \gamma = 2.0 \times w 
\label{eqn:Alpha}
\end{equation}
with $x^w = \lambda = \sin \theta _C$, 
where $\theta _C $ is the Cabbibo angle. 
In this parametrization we get 
\begin{equation} 
   V^{CKM} = \left(
      \begin{array}{ccc}
        1 - O(\lambda ^2)  &  O(\lambda )  &  
                              O(\lambda ^3)   \\ 
        O(\lambda )  &  1 - O(\lambda ^2)  &  
                              O(\lambda ^2)   \\ 
        O(\lambda ^3)  &   O(\lambda ^2) & 
                              1 - O(\lambda ^4)    \\
      \end{array} 
             \right).
\label{eqn:CKMN} 
\end{equation} 
Further, if we set 
\begin{equation}
    \beta = 2.5 \times w, \qquad 
    \delta = 1.5 \times w, 
\label{eqn:Beta}
\end{equation}
then we have quark masses 
\begin{eqnarray}
    m_u & = & O(\lambda ^7 v_u), \qquad 
      m_c = O(\lambda ^{3.5} v_u), \qquad 
        m_t = O(v_u), \\ 
    m_d & = & O(\lambda ^7 v_d), \qquad 
      m_s = O(\lambda ^6 v_d), \qquad 
        m_b = O(\lambda ^3 v_d). 
\end{eqnarray}
These results are in line with the observed values. 
Since $\alpha $, $\beta $, $\gamma $, and $\delta $ 
are set to be even positive integers, 
$w$ should be a multiple of 4 in this case. 
Taking $x^w = \lambda \sim 0.22$ and 
$x^{2sk-2} = m_{3/2}/M_S = 10^{-(15 \sim 16)}$ 
into account, 
we obtain the constraint 
\begin{equation}
     sk = (10 \sim 14) \times w. 
\label{eqn:sk}
\end{equation}
As a typical example, we will often refer 
the set of parameters 
\begin{eqnarray}
    & & s = 10, \qquad k = 5, \qquad 
        w = 4, \qquad e = 0,      \nonumber   \\
    & & \alpha = 4, \qquad \beta = 10, \qquad 
        \gamma = 8, \qquad \delta = 6, 
\label{eqn:nsk}
\end{eqnarray}
in which we have $K = 51$. 
In this case we have $x \simeq 0.7$ and 
$x^k \simeq 0.15$. 
Consequently, when $M_S = 10^{18}$GeV, 
the symmetry breaking scales $\langle S_0 \rangle$ 
and $\langle N^c_0 \rangle$ turn out to be 
$\sim 7 \times 10^{17}$GeV and 
$\sim 1.5 \times 10^{17}$GeV, respectively. 
Since the symmetry breaking scales are very 
large, 
we have the standard model gauge group 
over the wide energy range.


\setcounter{equation}{0}
\section{Spectra of leptons} 
\hspace*{\parindent} 
Let us now study the mass matrices for lepton sector, 
in which $L$-$H_d$ mixing occurs at energies 
below the scale $\langle N_0^c \rangle$. 
Colorless $SU(2)_L$-doublet fields $L$ and $H_d$ 
are not distinguished with each other 
under $G_{st}$. 
Then, hereafter we denote $R$-parity odd 
$H_{di}$ as $L'_i$ $(i=1,2,3)$. 
As mentioned in section 2, 
$H_{ui}$ and $L'_i$ $(i=1, 2, 3)$ 
in chiral multiplets do not 
develop their VEVs. 
It follows that there exist no mixings of 
$SU(2)_L \times U(1)_Y$ gauge superfields 
with $H_{ui}$ and $L'_i$ $(i=1, 2, 3)$. 
Since both $L$ and $L'$ are $SU(2)_L$-doublets, 
the CKM matrix for lepton sector becomes a unit matrix 
irrespective of the magnitude of $L$-$L'$ mixing. 
For charged leptons the superpotential is 
\begin{equation} 
   W_E \sim (S_0 \overline {S})^{h_{ij}} S_0 L'_i H_{uj} 
           + (S_0 \overline {S})^{m_{ij}} 
             L_i ( N^c_0 H_{uj} + H_{d0} E^c_j), 
\end{equation} 
where the exponents $h_{ij}$ are integers in the range 
$0 \leq h_{ij} < K$ and satisfy 
\begin{equation} 
    2 h_{ij} + b_i + b_j + a_0 + 2 \equiv 0  \qquad 
                           {\rm mod}\ K. 
\label{eqn:hbb}
\end{equation} 
Thus we have 
\begin{equation} 
   2 h_{ij} \equiv 2h_{33} + \left(
      \begin{array}{ccc}
        2\beta + 2\delta  &  \beta + 2\delta  &  
                              \beta + \delta    \\ 
        \beta + 2\delta  &  2\delta  &  \delta  \\ 
        \beta + \delta   &   \delta  &     0    \\
      \end{array} 
             \right)_{ij}      \qquad {\rm mod}\ K 
\label{eqn:hij} 
\end{equation} 
with $2h_{33} \equiv -2b_3 - a_0 - 2$. 
As before, we introduce a $3 \times 3$ matrix 
$H$ with elements 
\begin{equation} 
    H_{ij} = O(x^{2h_{ij}}). 
\end{equation} 
The mass matrix for charged leptons 
has the form 
\begin{equation} 
\begin{array}{r@{}l} 
   \vphantom{\bigg(}   &  \begin{array}{ccc} 
          \quad   H_u^+   &  \quad E^{c+}  &  
        \end{array}  \\ 
\widehat{M}_l = 
   \begin{array}{l} 
        L'^-  \\  L^-  \\ 
   \end{array} 
     & 
\left( 
  \begin{array}{cc} 
      x H    &    0       \\
      x^k M  &  \rho _d M 
  \end{array} 
\right) 
\end{array} 
\label{eqn:Mlh} 
\end{equation} 
in $M_S$ units. 
This $\widehat{M}_l$ is also a $6 \times 6$ 
matrix and can be diagonalized by 
a bi-unitary transformation as 
\begin{equation} 
    \widehat{\cal V}_l^{-1} \widehat{M}_l \, 
                        \widehat{\cal U}_l. 
\label{eqn:Ml} 
\end{equation} 
From Eqs.(\ref{eqn:Mlh}) and (\ref{eqn:Ml}) 
the matrix 
\begin{equation} 
    \widehat{\cal U}_l^{-1} \widehat{M}_l^{\dag } 
              \widehat{M}_l \, \widehat{\cal U}_l = 
    \widehat{\cal U}_l^{-1} 
      \left( 
      \begin{array}{cc}
          A_l + B_l    &  \epsilon _d B_l   \\
        \epsilon _d B_l   &  \epsilon _d^2 B_l 
      \end{array}
      \right)
    \widehat{\cal U}_l 
\end{equation} 
is diagonal, where 
\begin{equation} 
  A_l = x^2 H^{\dag} H, \qquad B_l = x^{2k} M^{\dag} M. 
\end{equation} 

The analysis is parallel to that of down-type 
quark masses in the previous section. 
We have the eigen equation 
\begin{equation} 
   \det \left( A_l + B_l - \frac {\eta }{M_S^2}
                \right) = 0 
\label{eqn:ABl} 
\end{equation} 
for heavy states. 
For three light states their masses 
squared are given by the eigen equation 
\begin{equation} 
   \det \left( x^{-2k}( A_l^{-1} + B_l^{-1} )^{-1}
         - \frac {\eta }{v_d^2} \right) = 0. 
\label{eqn:AB-l} 
\end{equation} 
The light states correspond to observed 
charged leptons. 
Due to $L$-$L'$ mixings 
mass pattern of charged leptons could be 
changed from that of up-type quarks. 
Introducing appropriate unitary matrices 
${\cal W}_l$ and ${\cal V}_l$, 
we can diagonalize $(A_l + B_l)$ and 
$(A_l^{-1} + B_l^{-1})^{-1}$ as 
\begin{equation} 
  {\cal W}_l^{-1} (A_l + B_l) {\cal W}_l 
                     = \Lambda _l^{(0)}, \qquad 
  {\cal V}_l^{-1} (A_l^{-1} + B_l^{-1})^{-1} 
                 {\cal V}_l = \Lambda _l^{(2)}, 
\end{equation} 
where $\Lambda _l^{(0)}$ and $\Lambda _l^{(2)}$ 
are diagonal $3 \times 3$ matrices. 
Masses of charged leptons are written as 
\begin{equation} 
   m_{li}^2 = v_d^2 \left( x^{-2k} 
            \Lambda _l^{(2)} \right)_{ii} 
                   \qquad (i = 1, 2, 3). 
\end{equation} 
Explicit forms of ${\widehat{\cal V}}_l$ 
and ${\widehat{\cal U}}_l$ are 
\begin{eqnarray} 
   \widehat{\cal V}_l & \simeq & \left( 
   \begin{array}{cc} 
      xH {\cal W}_l \,(\Lambda _l^{(0)})^{-1/2}  
         &  -(xH^{\dag})^{-1} {\cal V}_l 
                     \,(\Lambda _l^{(2)})^{1/2}  \\
      x^k M {\cal W}_l \,(\Lambda _l^{(0)})^{-1/2}   
         &  (x^kM^{\dag})^{-1} {\cal V}_l 
                         \,(\Lambda _l^{(2)})^{1/2} \\
   \end{array} 
                       \right), 
\label{eqn:hatvl}                              \\
   \widehat{\cal U}_l & \simeq & \left( 
   \begin{array}{cc} 
      {\cal W}_l   &  -\epsilon _d (A_l + B_l)^{-1} 
                                     B_l {\cal V}_l \\
      \epsilon _d B_l (A_l + B_l)^{-1} {\cal W}_l    &  
                                           {\cal V}_l 
   \end{array} 
                       \right).  
\end{eqnarray} 
In the same way as the case of down-type quarks, 
we also choose a large mixing solution. 
Defining an even integer $\xi$ by 
\begin{equation}
       \xi = 2h_{33}-k+1-(\alpha + \gamma 
              - \beta - \delta ), 
\end{equation}
we now impose the condition 
\begin{equation}
     \alpha + \gamma + \xi 
               - \beta - \delta \sim 0. 
\end{equation}
This condition means that 
\begin{equation}
    2h_{33} \sim k-1. 
\end{equation}
In this case eigenvalues of $A_l + B_l$ 
become $x^{2(k+\beta +2\delta )}$, 
$x^{2(k+\delta )}$ and $x^{2k}$. 
Thus masses of three heavy charged leptons 
are 
\begin{equation}
   M_S\, x^{k+\beta+2\delta }, \qquad 
   M_S\, x^{k+\delta }, \qquad 
   M_S\, x^{k }, 
\end{equation}
whose eigenstates are mainly 
$L_1$-$H_{u1}$, $L'_1$-$H_{u2}$ 
and $(O(1)L'_2+O(1)L'_3)$-$H_{u3}$, 
respectively. 
On the other hand, light charged leptons have 
their masses of 
\begin{equation}
   v_d\, x^{\alpha + \beta + \gamma + \delta + \xi }, 
                                         \qquad 
   v_d\, x^{\alpha + \gamma + \delta }, \qquad 
   v_d\, x^{\gamma }, 
\label{eqn:lmass}
\end{equation}
which correspond to observed $e$-, $\mu $- 
and $\tau $-leptons, respectively. 
Main components of these eigenstates are 
$L_2$-$E^c_1$, $L_3$-$E^c_2$ 
and $(O(1)L'_2+O(1)L'_3)$-$E^c_3$, 
respectively. 
The unitary matrix ${\cal V}_l$ 
which diagonalizes $A_l^{-1} + B_l^{-1}$, 
is expressed as 
\begin{equation}
    {\cal V}_l  =  
      \left( 
      \begin{array}{ccc}
        1 - O(x^{2\beta })  &  O(x^{\beta})  
                          &  O(x^{\beta + \delta })  \\
        O(x^{\beta })    &  1 - O(x^{2\delta })  
                                   &  O(x^{\delta })  \\
        O(x^{\beta + \delta })   &  O(x^{\delta })  
                              &  1 - O(x^{2\delta })  \\
      \end{array}
      \right). 
\end{equation}
Taking 
\begin{equation}
     \xi = w
\label{eqn:xi}
\end{equation}
together with the above 
parametrization (\ref{eqn:Alpha}) and (\ref{eqn:Beta}), 
we obtain 
\begin{equation}
    m_e  =  O(\lambda ^8 v_d), \qquad 
      m_{\mu } = O(\lambda ^{4.5} v_d), \qquad 
        m_{\tau } = O(\lambda ^2 v_d) 
\end{equation}
and 
\begin{equation} 
   {\cal V}_l = \left(
      \begin{array}{ccc}
        1 - O(\lambda ^5)  &  O(\lambda ^{2.5})  &  
                              O(\lambda ^4)   \\ 
        O(\lambda ^{2.5})  &  1 - O(\lambda ^3)  &  
                              O(\lambda ^{1.5})   \\ 
        O(\lambda ^4)  &   O(\lambda ^{1.5}) & 
                              1 - O(\lambda ^3)    \\
      \end{array} 
             \right).
\end{equation} 

We now proceed to study mass pattern of 
neutral sector. 
In the present framework there are fifteen 
neutral fields, i.e., $H^0_{ui}$, $L'^0_i$, 
$L^0_i$, $N^c_i$ and $S_i$ $(i=1,\,2,\,3)$. 
For neutral fields the superpotential is 
of the form 
\begin{eqnarray} 
   W_N & \sim & (S_0 \overline {S})^{h_{ij}} 
                              S_0 L'_i H_{uj} 
           + (S_0 \overline S)^{m_{ij}} 
             L_i ( N^c_0 H_{uj} + H_{u0} N^c_j) 
                                        \nonumber  \\
       & & \quad + \ (S_0 \overline S)^{s_{ij}} 
                (S_i \overline S) (S_j \overline S) 
             + (S_0 \overline S)^{t_{ij}} 
                (S_i \overline S) 
                 (N^c_j {\overline N^c})  \nonumber \\
       & & \quad + \ (S_0 \overline S)^{n_{ij}} 
                       (N^c_i {\overline N^c}) 
                          (N^c_j {\overline N^c}), 
\end{eqnarray} 
where the exponents $s_{ij}$, $t_{ij}$ and $n_{ij}$ 
are determined by 
\begin{eqnarray} 
  & & 2s_{ij}  + a_i + a_j + 2{\overline a} 
                   + 2 \equiv 0,      \nonumber \\
  & & 2t_{ij}  + a_i + b_j + {\overline a} 
                   + {\overline b} + 2 \equiv 0, 
                       \qquad {\rm mod}\ K  \\
  & & 2n_{ij}  + b_i + b_j + 2{\overline b} 
                   + 2 \equiv 0       \nonumber 
\end{eqnarray} 
in the range $0 \leq s_{ij}, t_{ij}, n_{ij} < K$. 
From Eqs.(\ref{eqn:mab}), (\ref{eqn:zaa}) and 
(\ref{eqn:hbb}) these equations are put 
into the form 
\begin{eqnarray} 
  & & 2s_{ij}  \equiv 2z_{ij} + a_0 - 2{\overline a}, 
                                          \nonumber \\
  & & 2t_{ij}  \equiv 2m_{ij} + b_0 - {\overline a} 
              - {\overline b}, \qquad {\rm mod}\ K  \\
  & & 2n_{ij}  \equiv 2h_{ij} + a_0 - 2{\overline b}. 
                                     \nonumber 
\end{eqnarray} 
In the above we have imposed the ansatzs 
\begin{eqnarray}
      2m_{33} & \equiv & 0, \\
      2z_{33} & \equiv & k-1 - \alpha - \gamma 
                              + \beta +\delta , \\
      2h_{33} & \equiv & k-1 + \alpha + \gamma 
                  - \beta -\delta + \xi  \sim k-1 
\end{eqnarray}
together with the conditions 
\begin{equation}
   a_0 + {\overline a} \equiv 2, \qquad 
   b_0 + {\overline b} \equiv 2k, \qquad 
   a_0 + 2b_0 \equiv 2e, \qquad 
   e = 0, 1. 
\end{equation}
Therefore, the exponents $s_{ij}$, $t_{ij}$ 
and $n_{ij}$ are rewritten as 
\begin{eqnarray} 
  & & 2s_{ij}  \equiv 2z_{ij} - 2 + 2e - 2k - \xi , 
                                          \nonumber \\
  & & 2t_{ij}  \equiv 2m_{ij} - 2 + 2e - 2k, 
                            \qquad  \qquad {\rm mod}\ K  \\
  & & 2n_{ij}  \equiv 2h_{ij} + 2e - 4k.   \nonumber 
\end{eqnarray} 

Introducing $3 \times 3$ matrices $S$, $T$ and $N$ 
with elements 
\begin{equation} 
     S_{ij} = O(x^{2s_{ij}}), \quad 
     T_{ij} = O(x^{2t_{ij}}), \quad 
     N_{ij} = O(x^{2n_{ij}}), 
\end{equation} 
we have a $15 \times 15$ mass matrix 
\begin{equation} 
\begin{array}{r@{}l} 
   \vphantom{\bigg(}   &  \begin{array}{cccccc} 
          \quad \, H_u^0   &  \ \  L'^0  &  \quad L^0  
                          &  \quad N^c   & \quad \quad  S  &
        \end{array}  \\ 
\widehat{M}_N = 
   \begin{array}{l} 
        H_u^0  \\  L'^0  \\  L^0  \\  N^c  \\  S  \\
   \end{array} 
     & 
\left( 
  \begin{array}{ccccc} 
      0     &  x H  &  x^k M^T   &     0     &    0   \\
      x H   &   0   &    0       &     0     &    0   \\
      x^k M &   0   &    0       & \rho _u M &    0   \\
      0     &   0   & \rho _u M^T 
                            & x^{2k} N  & x^{k+1} T^T \\
      0     &   0   &    0       & x^{k+1} T & x^2 S  \\
  \end{array} 
\right) 
\end{array} 
\label{eqn:Mn} 
\end{equation} 
in $M_S$ units for neutral sector, 
where $\rho _u = v_u/M_S$. 
Since $SU(2)_L$ symmetry is preserved above the 
electroweak scale, 
the eigen equation for six heavy states is 
the same as Eq.(\ref{eqn:ABl}). 
For another nine states 
we have an approximate eigen equation 
\begin{equation} 
   \det \left( \widehat{M}_{LNS} - 
               \frac {\eta }{M_S} \right) = 0, 
\label{eqn:LNS} 
\end{equation} 
where $\widehat{M}_{LNS}$ is defined by 
\begin{equation} 
   \widehat{M}_{LNS} = \left( 
   \begin{array}{ccc} 
       0           &  
         \epsilon _u (\Lambda ^{(2)}_l)^{1/2}{\cal V}_l^{-1}  
            &  0           \\
       \epsilon _u{\cal V}_l(\Lambda ^{(2)}_l)^{1/2} 
              &  x^{2k} N    &  x^{k+1} T^T \\
       0           &  x^{k+1} T   &  x^2 S   
   \end{array} 
   \right) 
\label{eqn:MLNS}
\end{equation} 
with $\epsilon _u = x^{-k}\rho _u$. 
Light neutrino masses are given by 
\begin{equation} 
   m_{\nu i} = \frac {m_{li}^2}{M_S \,x^{2k}}
             \left( \frac {v_u}{v_d} \right)^2 
             \left( {\cal V}_l^{-1} \Delta _N 
                        {\cal V}_l \right)_{ii} 
\label{eqn:Majo} 
\end{equation} 
with 
\begin{equation} 
   \Delta _N = (N-T^T S^{-1} T )^{-1}. 
\end{equation} 
This type of mass matrix has been discussed 
in Ref.\cite{Double}. 
Let us suppose that $\widehat{M}_{LNS}$ is a 
$3 \times 3$ matrix. 
When $N -TS^{-1}T \sim N$ in order of magnitude, 
the usual seesaw mechanism
\cite{Seesaw} 
is at work. 
On the other hand, when $N < TS^{-1}T$, 
another type of seesaw mechanism takes place. 
In the present framework $\widehat{M}_{LNS}$ 
is a $9 \times 9$ matrix. 
The exponents $2n_{ij}$ in $N_{ij}$ 
are equal to those in $(T^TS^{-1}T)_{ij}$ 
in modulus $K$. 
The relative magnitude of $N_{ij}$ to 
$(T^TS^{-1}T)_{ij}$ depends on the value of $k$. 
In the case $k > \frac {1}{2}(\alpha + \gamma ) 
- 1 + e$ which corresponds to small $s$ 
($s \lsim 8$), 
light neutrino masses are not so small. 
For instance, we have $m_{\nu \tau} = O$(1keV). 
So we do not adopt this case. 
In the case $k \leq \frac {1}{2}(\alpha + \gamma ) 
- 1 + e$ which corresponds to $s \gsim 8$, 
light neutrino masses are extremely small. 
Specifically, the latter case yields 
\begin{equation}
     \left( {\cal V}_l^{-1} \Delta _N 
                        {\cal V}_l \right)_{ii} 
     = \{x^{3k+1-2(\beta +\delta +e)}, \quad 
          x^{3k+1-2(\delta +e)}, \quad x^{3k+1-2e} \}_i. 
\end{equation}
Combining this with the result (\ref{eqn:lmass}) 
for $m_{li}$, 
we have the light neutrino masses 
\begin{equation}
     m_{\nu i} = \frac {v_u^2}{M_S} x^{k+1-2e} 
           \times \{x^{2(\alpha + \gamma + \xi )}, \quad 
               x^{2(\alpha + \gamma )}, \quad x^{2\gamma} \}. 
\end{equation}
The previous parametrization (\ref{eqn:Alpha}) 
and (\ref{eqn:Beta}) leads us to 
\begin{equation}
     m_{\nu i} = \frac {v_u^2}{M_S} x^{k+1-2e} 
              \times \{\lambda ^8, \quad \lambda ^6, 
                        \quad \lambda ^4 \}. 
\end{equation}
It follows that 
\begin{equation}
      \frac {m_{\nu e}}{\lambda ^4} \sim 
        \frac {m_{\nu \mu}}{\lambda ^2} \sim 
          m_{\nu \tau} \leq O(10^{-7}{\rm eV}). 
\end{equation}
The calculated neutrino masses seem to be 
too small. 
According to the analyses of solar neutrino
\cite{Solar}, atmospheric neutrino\cite{Atmos} 
and cosmological constraints\cite{Cosmo}, 
it is preferable that three typical mass scales 
of neutrinos are $\sim 10^{-3}$eV, $\sim 10^{-1}$eV 
and $\sim 10$eV. 
The ratios $m_{\nu e}/m_{\nu \mu }$ and 
$m_{\nu \mu}/m_{\nu \tau }$ obtained here 
are consistent with those among the above 
three typical mass scales. 
As pointed out at the beginning of this section, 
the CKM-matrix for lepton sector is a unit 
matrix irrespective of the magnitude of 
$L$-$L'$ mixing. 
The situation is unchanged even through 
seesaw mechanism. 
This is because the fields $N^c$ and $S$ are 
$SU(2)_L$-singlet. 
In addition, the components of $N^c$ and $S$ 
in light neutrinos are very small. 
Thus we have no flavor-changing charged currents 
at tree level. 
Recently, by introducing discrete symmetries 
which yield appropriate texture zeros in 
Yukawa couplings, 
the hierarchical pattern of neutrino 
masses has been examined in Ref.\cite{Suema}. 
Finally, we touch upon the remaining eigenvalues 
of Eq.(\ref{eqn:LNS}). 
Three pairs of heavy states which are 
$G_{st}-$neutral have their masses of 
\begin{equation}
   M_S x^{-k-1+2e} \times 
      \{x^{\beta + \delta - \alpha - \xi }, \quad 
          x^{\alpha + \gamma }, \quad 
            x^{\alpha + \gamma + \delta + \xi } \}. 
\end{equation}


\setcounter{equation}{0}
\section{Proton decay} 
\hspace*{\parindent} 
After studying the particle spectra of down-type 
colored fields in $\Phi _0$ and ${\overline \Phi }$, 
we explore the proton stability in this section. 
Under the discrete $R$-symmetry 
the superpotential of down-type colored fields is 
of the form 
\begin{eqnarray} 
     W_g & \sim & (S_0 \overline S)^q S_0 g_0 g^c_0 + 
                (S_0 \overline S)^{sk-4-q} 
             (\overline S \overline g \,{\overline g^c} 
             + (S_0 \overline S)^{1-e} 
                N^c_0 \overline S D^c_0 {\overline g^c}) 
                                     \nonumber  \\
       & &  + \ (S_0 \overline S)^{sk-1} (g_0 \overline g 
             +  g^c_0 {\overline g^c} 
             +  (S_0 \overline S)^{1-e} g_0 N^c_0 D^c_0 ) 
             + (S_0 \overline S)^{(s-2)k+e-1} 
             \overline g {\overline N^c} {\overline D^c} 
                                     \nonumber  \\
       & &  + \ (S_0 \overline S)^{(s-1)k} 
                                D^c_0 {\overline D^c} 
             + (S_0 \overline S)^{(s-2)k+q+e+1} 
             S_0 {\overline N^c} g^c_0 {\overline D^c}, 
\end{eqnarray} 
where $q = k + \frac {1}{2}\xi - e - 2$. 
When $S_0$, $\overline S$, $N^c_0$ and ${\overline N^c}$ 
develop non-zero VEVs, 
the superpotential induces the mass matrix 
\begin{equation} 
\begin{array}{r@{}l} 
   \vphantom{\bigg(}    &  \begin{array}{cccc} 
            \qquad \qquad  g^c_0 \qquad \qquad 
                & \quad  \overline{g} 
                         \qquad \qquad & 
                           \qquad  D^c_0  &  
         \end{array}  \\
   M_g = 
     \begin{array}{l} 
        g_0  \\  {\overline D^c}  \\  {\overline g^c} \\
     \end{array} 
      & 
\left(
  \begin{array}{ccc}
  O(x^{2k-3-2e+\xi })          &  O(\rho )    
                           &  O(\rho \,x^{k+2-2e})    \\ 
  O(\rho \,x^{1-k+\xi }) & O(\rho \,x^{-3k+2e}) 
                           & O(\rho \,x^{-2k+2}) \\ 
  O(\rho )           &  O(\rho \,x^{-2k-1+2e-\xi }) 
                           &  O(\rho \,x^{-k+1-\xi }) 
  \end{array}
\right) 
\end{array} 
\end{equation} 
in $M_S$ units, 
where $\rho = m_{3/2}/M_S = 10^{-(15 \sim 16)}$. 
This mass matrix is diagonalized by a bi-unitary 
transformation as 
\begin{equation}
    {\cal V}_g^{-1} M_g \, {\cal U}_g 
\end{equation}
with 
\begin{eqnarray} 
    {\cal V}_g & = & 
      \left( 
      \begin{array}{ccc}
        1   &  O(\rho \,x^{-3k+4+2e})  
                          &  O(\rho \,x^{-2k+3+2e-\xi })  \\
        O(\rho \,x^{-3k+4+2e})    &  1  
                                   &  O(x^{k-1-\xi })  \\
        O(\rho \,x^{-2k+3+2e-\xi })   &  O(x^{k-1-\xi })  
                              &  1   \\
      \end{array}
      \right), \\
    {\cal U}_g & = & 
      \left( 
      \begin{array}{ccc}
        1   &  O(\rho \,x^{-2k+3+2e})  
                          &  O(\rho \,x^{-k+5-\xi })  \\
        O(\rho \,x^{-2k+3+2e})    &  1  
                                   &  O(x^{k+2-2e})  \\
        O(\rho \,x^{-k+5-\xi })   &  O(x^{k+2-2e})  
                              &  1   \\
      \end{array}
      \right). 
\end{eqnarray} 
The eigenvalues are given by 
\begin{eqnarray}
   M_{gA} & = & O(M_S x^{2k-3-2e+\xi }), \nonumber \\ 
   M_{gB} & = & O(m_{3/2} x^{-3k+2e}),   \\ 
   M_{gC} & = & O(m_{3/2} x^{-k+1-\xi }). \nonumber 
\label{eqn:gmass}
\end{eqnarray}
From explicit forms of ${\cal V}_g$ and 
${\cal U}_g$ we find that 
three eigenstates $g_A$-$g^c_A$, 
$g_B$-$g^c_B$ and $g_C$-$g^c_C$ are approximately 
$g_0$-$g_0^c$, ${\overline D^c}$-${\overline g}$ 
and ${\overline g^c}$-$D_0^c$ states, respectively. 
In the typical example (\ref{eqn:nsk}) 
$M_{gA}$ is nearly GUT-scale ($\sim 10^{16}$GeV). 
By contrast, $M_{gB}$ and $M_{gC}$ are as 
small as $O(10^{5 \sim 6}$GeV). 
Then, at first sight, it seems that dimension-five 
operators mediated by these rather light colored 
fields lead to fast proton decay. 
However, this is not the case. 
The dimension-five operators mediated 
by light colored fields are strongly suppressed 
because of extremely small effective couplings. 
This is due to the fact that 
(1,2), (2,1), (1,3) and (3,1) entries of 
${\cal V}_g$ are $O(10^{-13})$. 
In what follows we explain this situation 
more explicitly.

Since $R$-parity of quark and lepton superfields 
are odd, 
those of $SU(2)_L$-singlet colored superfields 
mediating proton decay should be even. 
The relevant superfields are $g_0$, $g_0^c$, 
and $D_0^c$ which reside in $\Phi _0({\bf 27})$. 
Effective trilinear couplings with $g_0$, $g_0^c$ 
and $D_0^c$ are given by 
\begin{eqnarray}
    W_g^{eff} & \sim & (Q^TZQ + N^{cT}HD^c 
                              + E^{cT}HU^c)g_0 
                + (Q^TZL + U^{cT}HD^c)g_0^c    \nonumber \\
              &  &  \qquad + (Q^TML' + D'^TMN^c 
                           + D'^{cT}MU^c)D_0^c, 
\end{eqnarray}
where $3 \times 3$ matrices $Z$, $H$ and $M$ 
have already been determined in sections 4 and 5. 
Superfields $g_0$, $g_0^c$ and $D_0^c$ are expressed 
in terms of mass eigenstates as 
\begin{eqnarray}
    g_0   & = & {\cal V}_{g11} \,g_A 
                + {\cal V}_{g12} \,g_B 
                  + {\cal V}_{g13} \,g_C,  \nonumber  \\
    g_0^c & = & {\cal U}_{g11} \,g^c_A 
                + {\cal U}_{g12} \,g^c_B 
                  + {\cal U}_{g13} \,g^c_C,        \\
    D_0^c & = & {\cal U}_{g31} \,g^c_A 
                + {\cal U}_{g32} \,g^c_B 
                  + {\cal U}_{g33} \,g^c_C.    \nonumber 
\end{eqnarray}
Taking ${\cal V}_{g11},\ {\cal U}_{g11} \simeq 
1$ and ${\cal U}_{g31} \sim \rho x^{-k+5-\xi }$ 
into account, 
we can obtain dominant dimension-five operators 
from $g_A$-$g^c_A$ exchange 
\begin{equation}
   \frac {1}{M_S} x^{-2k+3+2e-\xi} 
     (Q^TZQ + N^{cT}HD^c + E^{cT}HU^c)
            (Q^TZL + U^{cT}HD^c). 
\label{eqn:opr1}
\end{equation}
Similarly, dominant dimension-five operators from 
$g_C$-$g^c_C$ exchange are 
\begin{equation}
   \frac {1}{M_S} x^{-k+2+2e} 
     (Q^TZQ + N^{cT}HD^c + E^{cT}HU^c)
         (Q^TML' + D'^TMN^c + D'^{cT}MU^c). 
\label{eqn:opr2}
\end{equation}
The prefactor is induced from 
${\cal V}_{g13}(M_{gC})^{-1}{\cal U}_{g33}$. 
In dimension-five operators for 
$g_B$-$g^c_B$ exchange the prefactor 
is ${\cal V}_{g12}(M_{g_B})^{-1}{\cal U}_{g32}$, 
which is smaller than the one for 
$g_C$-$g^c_C$ exchange 
by the factor $x^{2k+4(1-e)}$. 
Therefore, the study of dimension-five operators 
coming from $g_A$-$g^c_A$ and 
$g_C$-$g^c_C$ exchanges 
suffices to explore the proton stability.

We now rewrite the above operators in terms of 
quark and lepton mass eigenstates, 
which are represented by using the symbol 
"tilde". 
In order to implement this rewriting, 
we can use the transfer 
\begin{eqnarray}
   Q   & \rightarrow &   {\cal V}_u \tilde{Q}'
        = \left(
           \begin{array}{cc}
            {\cal V}_u  &     0       \\ 
               0        &  {\cal V}_u 
           \end{array}
          \right)
          \left( 
           \begin{array}{c}
                 \tilde{U}     \\
             V^{CKM} \tilde{D}
           \end{array}
          \right),    
                                  \nonumber   \\
   U^c & \rightarrow &   {\cal U}_u \tilde{U^c}, 
        \qquad  D^c  \rightarrow   
              \widehat{\cal U}_{d22} \tilde{D^c}, 
                                  \nonumber   \\ 
   D'  & \rightarrow &   {\cal V}_d \tilde{D}, 
        \qquad \  D'^c  \rightarrow   
              \widehat{\cal U}_{d12} \tilde{D^c}, 
                                              \\
   L   & \rightarrow &   \widehat{\cal V}_{l22} 
                                      \tilde{L}, 
        \qquad  L'  \rightarrow   
              \widehat{\cal V}_{l12} \tilde{L}, 
                                  \nonumber    \\ 
   E^c & \rightarrow &   {\cal V}_l \tilde{E^c}, 
                                     \nonumber 
\end{eqnarray}
where 
\begin{eqnarray}
     \widehat{\cal U}_{d12} & = & 
                   -(x Z)^{-1} {\cal V}_d 
                          (\Lambda _d^{(2)})^{1/2},   \\ 
     \widehat{\cal U}_{d22} & = & 
                    (x^k M)^{-1} {\cal V}_d 
                          (\Lambda _d^{(2)})^{1/2},   \\ 
     \widehat{\cal V}_{l12} & = & 
                   -(x H^{\dag})^{-1} {\cal V}_l 
                          (\Lambda _l^{(2)})^{1/2},   \\ 
     \widehat{\cal V}_{l22} & = & 
                    (x^k M^{\dag})^{-1} {\cal V}_l 
                          (\Lambda _l^{(2)})^{1/2}, 
\end{eqnarray}
which are three-by-three blocks of matrices given in 
Eqs.(\ref{eqn:hatud}) and (\ref{eqn:hatvl}). 
Light component of $N^c$ is extremely small 
and then its contribution to nucleon decay is 
negligible. 
Therefore, Eq. (\ref{eqn:opr1}) is translated into 
\begin{eqnarray}
    \lefteqn{ \frac{1}{M_S} x^{-2k+3+2e-\xi }  
       [ \tilde{Q}'^T({\cal V}_u^TZ{\cal V}_u)
         \tilde{Q}'\times \tilde{Q}'^T({\cal V}_u^T Z
           \widehat{\cal V}_{l22})\tilde{L} }  
                                  \nonumber  \\
    & & \hphantom{QQQQ} 
        + \,\tilde{E}^{cT}({\cal V}_l^TH{\cal U}_u)
         \tilde{U}^c \times \tilde{U}^{cT}({\cal U}_u^T H
           \widehat{\cal U}_{d22})\tilde{D}^c ]. 
\label{eqn:oprA}
\end{eqnarray}
Similarly, Eq. (\ref{eqn:opr2}) is put into 
\begin{eqnarray}
    \lefteqn{ \frac{1}{M_S} x^{-k+2+2e}  
       [ \tilde{Q}'^T({\cal V}_u^TZ{\cal V}_u)
         \tilde{Q}'\times \tilde{Q}'^T({\cal V}_u^T M
           \widehat{\cal V}_{l12})\tilde{L} }  
                                   \nonumber  \\
    & & \hphantom{QQQQ} 
        + \,\tilde{E}^{cT}({\cal V}_l^TH{\cal U}_u)
         \tilde{U}^c \times \tilde{U}^{cT}({\cal U}_u^T M^T
           \widehat{\cal U}_{d12})\tilde{D}^c ]. 
\label{eqn:oprC}
\end{eqnarray}
The dimension-five operators result in 
nucleon decay via gaugino- or Higgsino-dressing 
processes
\cite{Pdecay}\cite{Pdcal}. 
Among various dressing processes 
the exchange of charged wino or Higgsino 
give predominant contributions to nucleon decay. 
Since $SU(2)_L$-singlet states 
$\tilde{U^c}$, $\tilde{D^c}$, $\tilde{E^c}$ 
do not couple to $SU(2)_L$-gauginos, 
dominant dimension-five operators with charged 
wino-dressing processes turn out to be 
the first terms in 
Eqs.(\ref{eqn:oprA}) and (\ref{eqn:oprC}). 
Thus we have dominant operators incorporating 
charged wino-dressing processes 
\begin{eqnarray}
   \lefteqn{ \frac {1}{M_S} x^{-2k+3+2e-\xi }
     \tilde{Q'_1}({\cal V}_u^T Z {\cal V}_u)_{11}
       \tilde{Q'_1} \times \tilde{Q'_2}({\cal V}_u^T 
          Z \widehat{\cal V}_{l22})_{2j}\tilde{L_j} } 
                                   \nonumber  \\
   & & + \frac {1}{M_S} x^{-k+2+2e} 
     \tilde{Q'_1}({\cal V}_u^T Z {\cal V}_u)_{11}
       \tilde{Q'_1} \times \tilde{Q'_2}({\cal V}_u^T 
          M \widehat{\cal V}_{l12})_{2j}\tilde{L_j}. 
\label{eqn:QQQL}
\end{eqnarray}
Simple calculations yield 
\begin{eqnarray}
    ({\cal V}_u^T Z {\cal V}_u)_{ij} & = & O(Z_{ij}), 
                                                     \\
    ({\cal V}_u^T Z \widehat{\cal V}_{l22})_{ij} & = & 
                              O(x^{k-1+\xi }) 
       ({\cal V}_u^T M \widehat{\cal V}_{l12})_{ij}  
                                     \nonumber  \\
     & = & x^{k-1+\beta + \delta } \times 
     \left(
     \begin{array}{ccc}
     O(x^{\alpha + \gamma + \xi }) & 
          O(x^{\alpha + \gamma })  &  O(x^{\gamma }) \\ 
     O(x^{\gamma + \xi })  &  O(x^{\gamma })  &  
                             O(x^{\gamma - \alpha }) \\ 
     O(x^{\xi })  &  O(1)  &  O(x^{-\alpha })
     \end{array}
     \right)_{ij}. 
\end{eqnarray}
From these relations Eq.(\ref{eqn:QQQL}) becomes 
\begin{equation}
    \frac {1}{M_S} (\tilde{Q'_1} \tilde{Q'_1}
         \tilde{Q'_2} \tilde{L_j}) \, 
         x^{2(\alpha + \beta + \gamma + \delta )
             +1+2e-\xi } \times 
             (x^{-\alpha + \xi }, \ x^{-\alpha }, 
                \ x^{-2\alpha })_j. 
\end{equation}
This implies that a dominant mode of proton decay 
is $p \rightarrow K^0\ + \mu ^+$. 
In this decay mode the magnitude of the dimension-five 
operator is given by 
\begin{equation}
     \frac {1}{M_S} x^{\alpha + 2(\beta + \gamma + \delta )
                                     +1+2e-\xi } 
       \  \simeq \  \frac {1}{M_S} \lambda ^{12}
           \ \gsim \ 10^{-(25.5 \sim 26.5)} {\rm GeV}^{-1}. 
\end{equation}

The second terms in Eqs.(\ref{eqn:oprA}) and 
(\ref{eqn:oprC}) contribute to nucleon decay via 
charged Higgsino-dressing processes. 
The relevant terms are 
\begin{eqnarray}
   \lefteqn{ \frac {1}{M_S} x^{-2k+3+2e-\xi }
     \tilde{E}^c_i({\cal V}_l^T H {\cal U}_u)_{ij}
       \tilde{U}^c_j \times \tilde{U}^c_m({\cal U}_u^T 
          H \widehat{\cal U}_{d22})_{m1}\tilde{D}^c_1 } 
                                  \nonumber  \\
   & & + \frac {1}{M_S} x^{-k+2+2e} 
     \tilde{E}^c_i({\cal V}_l^T H {\cal U}_u)_{ij}
       \tilde{U}^c_j \times \tilde{U}^c_m({\cal U}_u^T 
          M \widehat{\cal U}_{d12})_{m1}\tilde{D}^c_1, 
\label{eqn:EUUD}
\end{eqnarray}
where $(j,\,m)=(1,\,2),\ (2,\,1)$. 
Using the relations 
\begin{eqnarray}
    ({\cal V}_l^T H {\cal U}_u)_{ij} & = & O(H_{ij}), 
                                                     \\
    ({\cal U}_u^T H \widehat{\cal U}_{d22})_{ij} & = & 
                              O(x^{k-1}) 
       ({\cal U}_u^T M \widehat{\cal U}_{d12})_{ij}  
                                  \nonumber  \\ 
     & = & x^{k-1+\gamma } \times 
     \left(
     \begin{array}{ccc}
     O(x^{\alpha + \beta + \delta }) & 
          O(x^{\alpha + \beta + \delta })  &  
              O(x^{\beta + \delta }) \\ 
     O(x^{\alpha + \delta })  &  
          O(x^{\alpha + \delta })  &  
              O(x^{\delta }) \\ 
     O(x^{\alpha })  &  O(x^{\alpha })  &  O(1)
     \end{array}
     \right)_{ij}, 
\end{eqnarray}
we obtain the dimension-five operators for 
$SU(2)_L$-singlet fields 
\begin{equation}
    \frac {1}{M_S} (\tilde{E^c_j} \tilde{U^c_2}
         \tilde{U^c_1} \tilde{D^c_1}) \,
         x^{\beta + 2(\alpha + \gamma + \delta )
             +1+2e } \times 
             (1, \ x^{-\beta }, \ x^{-\beta - \delta })_j. 
\end{equation}
In the Higgsino-dressing processes the operators 
are multiplied by their Yukawa couplings. 
As a consequence we have a dominant contribution 
in the case $j=3$. 
The above operator multiplied 
by the Yukawa couplings for $\tilde{E^c_3}$ 
and $\tilde{U^c_2}$ becomes 
\begin{equation}
     \frac {1}{M_S} x^{2(\alpha + 2\gamma + \delta )
                                     +1+2e} 
        \simeq  \frac {1}{M_S} \lambda ^{13}, 
\end{equation}
in magnitude and results in the decay 
$p \rightarrow K^+\ + {\overline \nu _{\tau}}$.

In conclusion of this section, 
the main mode of proton decay is 
$p \rightarrow K^0\ + \mu ^+$, 
in which the magnitude of the dimension-five 
operator is about $\lambda ^{12}/M_S \simeq 
10^{-(25.5 \sim 26.5)}{\rm Gev}^{-1}$. 
This implies that the proton lifetime 
is about $10^{33 \sim 35}$yr. 
This result is consistent with the present 
experimental data.


\setcounter{equation}{0}
\section{Gauge coupling unification} 
\hspace*{\parindent} 
As is well-known, there is a discrepancy between 
the string scale $M_S$ and the MSSM unification 
scale $\sim 2 \times 10^{16}$GeV. 
Main concern here is whether or not we can 
reconcile this discrepancy in the present model. 
For this purpose we study the renormalization group 
evolution of the gauge couplings in the model 
up to two-loop order.

In the preceding sections particle spectra have been 
already studied. 
Unlike the MSSM, in the present model there are 
many extra intermediate-scale fields, 
which are tabulated in Table II. 
In particular, the contributions of 
$(H_{u0},\ H_{d0},\ {\overline H_u},\ 
{\overline H_d})$ and $({\overline g},\ 
{\overline g^c},\ {\overline D^c},\ D^c_0)$ 
are significant, 
because their masses are lying in rather low 
energy region ($10^{2 \sim 7}$GeV).

\vspace {5mm} 
\begin{center} 
\framebox [3cm] {\large \bf Table II} 
\end{center} 
\vspace {5mm} 

The evolution equations for $\alpha _i = g_i^2/4\pi $ 
are generally given up to two-loop order by 
\begin{equation}
  \mu ^2 \frac {d\alpha _i}{d\mu ^2} = 
    \frac {1}{4\pi } \left[ \,b_i 
         + \sum _j \frac {b_{ij}}{4\pi } \alpha _j 
         - \frac {a_i}{4\pi }\, \right] \alpha _i^2, 
\end{equation}
where $\mu $ is the running mass scale
\cite{RGE}. 
The coefficients $b_i$, $b_{ij}$ and $a_i$ 
are determined by the particle content of the model. 
The third term in the r.h.s. represents 
the contribution of Yukawa couplings. 
In the present calculation we take accout only 
of the largest Yukawa couplings $f = M_{33}$, 
namely, Yukawa couplings of the third generation 
$\Phi _3({\bf 15,\,1})\Phi _3({\bf 6^*,\,2})
\Phi _0({\bf 6^*,\,2})$ 
and for simplicity we neglect the renormalization 
group evolution of the Yukawa couplings. 
In our analysis it is assumed that 
string threshold corrections are 
negligibly small.

In the region between $M_S$ and 
$\langle S_0 \rangle = M_S\,x$, 
where the gauge symmetry is $SU(6)\times SU(2)_R$, 
we have 
\begin{equation}
    b_i    = \left(
             \begin{array}{c}
                -8    \\
                 9 
             \end{array}
             \right),       \qquad 
    b_{ij} = \left(
             \begin{array}{cc}
                 9  &  15     \\
               175  &  81 
             \end{array}
             \right),       \qquad 
    \frac{a_i}{y}    = \left(
             \begin{array}{c}
                 28  \\
                 60 
             \end{array}
             \right), 
\end{equation}
where $y = f^2/4\pi $ is taken to be a constant. 
In the region between $M_S\,x$ and 
$\langle N_0^c \rangle = M_S\,x^k$, 
where the gauge symmetry is 
$SU(4)_{PS} \times SU(2)_L \times SU(2)_R$, 
we get 
\begin{equation}
    b_i    = \left(
             \begin{array}{c}
                 1    \\
                 5    \\
                 9 
             \end{array}
             \right),       \qquad 
    b_{ij} = \left(
             \begin{array}{ccc}
               118  &   9  &  15     \\
                45  &  53  &  15     \\
                75  &  15  &  81  
             \end{array}
             \right),       \qquad 
    \frac{a_i}{y}    = \left(
             \begin{array}{c}
                 36  \\
                 32+4n(S_3)  \\
                 56+4n(S_3) 
             \end{array}
             \right) 
\end{equation}
with 
\begin{eqnarray}
   n(S_3) & = & \left\{ 
             \begin{array}{ll}
             1  &  \ M_S\ x > \mu 
                   \geq  M_S\,x^{-k+\gamma+2e-1} \\
             0  &  \ M_S\ x^{-k+\gamma+2e-1} > \mu 
                            \geq  M_S\,x^k. 
             \end{array}
             \right.  
\end{eqnarray}
In the wide energy region ranging from 
$M_S\,x^k$ to $m_{3/2}=M_S\,x^{2sk-2}$ 
the gauge group coincides with the standard 
model gauge group. 
From Table II we can calculate the coefficients, 
which are of the forms 
\begin{eqnarray}
    b_i   & = & \left(
            \begin{array}{c}
              -3  \\
               0  \\
               6  
            \end{array}
            \right)
            + \left(
            \begin{array}{c}
               1  \\
               0  \\
               2/5  
            \end{array}
            \right) n_g 
            + \left( 
            \begin{array}{c}
               0  \\
               1  \\
               3/5  
            \end{array}
            \right) n_H ,  \\
    b_{ij} & = & \left(
            \begin{array}{ccc}
                 14     &      9      &  11/5   \\
                 24     &     18      &   6/5   \\
                88/5    &    18/5     &  38/5  
            \end{array}
            \right)
            + \left(
            \begin{array}{ccc}
                 34/3   &   0   &   4/15   \\
                   0    &   0   &    0     \\
                 32/15  &   0   &   8/75  
            \end{array}
            \right) n_g           \nonumber    \\
        & & \hphantom{QQQQ} + \left( 
            \begin{array}{ccc}
               0    &   0   &     0    \\
               0    &   7   &    3/5   \\
               0    &  9/5  &    9/25 
            \end{array}
            \right) n_H ,  \\
    \frac{a_i}{y}  & = & \left(
            \begin{array}{c}
               4  \\
               8  \\
               44/5  
            \end{array}
            \right) 
            + \left(
            \begin{array}{c}
               4  \\
               6  \\
               14/5  
            \end{array}
            \right) \,n(D^c_3) 
            + \left(
            \begin{array}{c}
               8  \\
               3  \\
               31/5  
            \end{array}
            \right) \,n(D^c_0)      \nonumber \\
        & &  \hphantom{QQQQ}  + \left( 
            \begin{array}{c}
               0  \\
               2  \\
               6/5  
            \end{array}
            \right) \,n(N^c_3).  
\end{eqnarray}
In these expressions $n_H$ and $n_g$ 
stand for the numbers of doublet Higgses and 
extra down-type colored fields, 
respectively and are given by 
\begin{eqnarray}
   n_H & = & \left\{ 
             \begin{array}{ll}
             4  &  \ M_S\,x^k > \mu 
                            \geq M_S\,x^{k+\delta } \\
             3  &  \ M_S\,x^{k+\delta } > \mu 
                            \geq  M_S\,x^{k+\beta +2\delta } \\
             2  &  \ M_S\,x^{k+\beta +2\delta } > \mu 
                            \geq  M_S\,x^{2sk-4k+2e-1} \\
             1  &  \ M_S\ x^{2sk-4k+2e-1} > \mu 
                            \geq  M_S\,x^{2sk-2}, 
             \end{array}
             \right.  \\
   n_g & = & \left\{ 
             \begin{array}{ll}
             5  &  \ M_S\,x^k > \mu 
                            \geq M_S\,x^{2k-3-2e+\xi } \\
             4  &  \ M_S\,x^{2k-3-2e+\xi } > \mu 
                    \geq  M_S\,x^{k+\beta+\delta-\alpha } \\
             3  &  \ M_S\,x^{k+\beta+\delta-\alpha } > \mu 
                  \geq  M_S\,x^{k+\alpha +\gamma+\delta } \\
             2  &  \ M_S\,x^{k+\alpha +\gamma+\delta } > \mu 
                            \geq  M_S\ x^{2sk-3k+2e-2} \\
             1  &  \ M_S\ x^{2sk-3k+2e-2} > \mu 
                            \geq  M_S\,x^{2sk-k-1-\xi } \\
             0  &  \ M_S\,x^{2sk-k-1-\xi } > \mu 
                            \geq  M_S\,x^{2sk-2}. 
             \end{array}
             \right.  
\end{eqnarray}
$n(D^c_3)$, $n(D^c_0)$ and $n(N^c_3)$ are 
\begin{eqnarray}
   n(D^c_3) & = & \left\{ 
             \begin{array}{ll}
             1  &  \ M_S\,x^k > \mu 
               \geq  M_S\,x^{k+\beta +\delta -\alpha } \\
             0  &  \ M_S\ x^{k+\beta +\delta -\alpha } > \mu 
                            \geq  M_S\,x^{2sk-2}, 
             \end{array}
             \right.  \\
   n(D^c_0) & = & \left\{ 
             \begin{array}{ll}
             1  &  \ M_S\,x^k > \mu 
               \geq  M_S\,x^{2sk-k-1+\xi } \\
             0  &  \ M_S\ x^{2sk-k-1+\xi } > \mu 
                            \geq  M_S\,x^{2sk-2}, 
             \end{array}
             \right.  \\
   n(N^c_3) & = & \left\{ 
             \begin{array}{ll}
             1  &  \ M_S\,x^k > \mu 
               \geq  M_S\,x^{\alpha +\gamma -k+2e-1} \\
             0  &  \ M_S\,x^{\alpha +\gamma -k+2e-1} > \mu 
                            \geq  M_S\,x^{2sk-2}. 
             \end{array}
             \right.  
\end{eqnarray}
It should be noted that in the present model 
we obtain $n_H - n_g = 0$ over rather wide energy range. 
By contrast, in the MSSM we have $n_H - n_g = 1$. 
In the region between $m_{3/2}$ and $M_Z$ 
where supersymmetry is broken, 
all superparticles except for light Higgses 
do not contribute the evolution equations. 
This leads us to 
\begin{equation}
    b_i    = \left(
             \begin{array}{c}
                     -7     \\
                    -7/3    \\
                    23/5 
             \end{array}
             \right),       \qquad 
    b_{ij} = \left(
             \begin{array}{ccc}
                  -26    &   9/2   &   11/10   \\
                   12    &  97/6   &    3/2    \\
                  44/5   &   9/2   &  217/50 
             \end{array}
             \right),       \qquad 
    \frac{a_i}{y}    = \left(
             \begin{array}{c}
                  2    \\
                  2    \\
                 16/5 
             \end{array}
             \right). 
\end{equation}

We are now in a position to solve 
the evolution equation numerically. 
The behavior of the renormalization group flow 
is shown in Fig.1, 
in which we choose a typical example (\ref{eqn:nsk}). 
In the present calculation the parameters 
are taken as 
\begin{equation}
    M_S = 0.5 \times 10^{18}{\rm GeV}, \quad 
    m_{3/2} = 200{\rm GeV}, \quad 
    \alpha _{string}^{-1} = 14.0, \quad 
    f = 1.7, 
\end{equation}
where $\alpha _{string}$ represents the unified 
gauge coupling at the string scale. 
Resulting values of $\alpha _i^{-1}(M_Z)$ 
are 
\begin{equation}
    \alpha _1^{-1}(M_Z) = 58.96, \quad 
    \alpha _2^{-1}(M_Z) = 26.03, \quad 
    \alpha _3^{-1}(M_Z) = 8.66.  
\end{equation}
Compared with the present experimental values 
\cite{PDG}
\begin{equation}
    \alpha _1^{-1}(M_Z) = 58.95 \pm 0.08, \quad 
    \alpha _2^{-1}(M_Z) = 29.66 \pm 0.07, \quad 
    \alpha _3^{-1}(M_Z) =  8.48 \pm 0.43,  
\end{equation}
the calculated $\alpha _1^{-1}(M_Z)$ and 
$\alpha _3^{-1}(M_Z)$ are consistent with the data, 
while the calculated $\alpha _2^{-1}(M_Z)$ 
is smaller than the observed one by $\sim 3.5$. 
Consequently, our analysis shows that 
the gauge coupling unification is not achieved 
at the string scale. 
This suggests that we are not successful in 
getting proper particle spectra of extra 
intermediate-scale matter.

\vspace {5mm} 
\begin{center} 
\framebox [3cm] {\large \bf Fig. 1} 
\end{center} 
\vspace {5mm} 


\setcounter{equation}{0}
\section{Summary} 
\hspace*{\parindent} 
In the context of level-one string model 
we have explored 
a possibility that characteristic pattern of 
quark-lepton masses and the CKM matrix have their 
origin in the discrete $R$-symmetry and mixing 
mechanism. 
In this paper we have chosen $Z_K \times Z_2$ 
symmetry with $K = sk + 1$ 
as an example of the discrete $R$-symmetry. 
The $Z_2$-symmetry is assumed so as to be in 
accord with the $R$-parity in the MSSM 
and is unbroken 
down to the electroweak scale. 
The vector-like multiplets $\Phi _0$, 
$\overline \Phi $ and the chiral multiplets 
$\Phi _i$ $(i=1, 2, 3)$ are assigned to 
even and odd $R$-parity, respectively. 
Under this assignment no mixing occurs between 
the vector-like multiplets and the chiral multiplets. 
The $Z_K$ symmetry is used as a horizontal 
symmetry. 
The $Z_K$ symmetry controls a large hierarchy 
of the energy scales of the symmetry breaking 
and particle spectra. 
Triplet-doublet splitting problem and 
the $\mu $-problem are solved as a 
result of the discrete symmetry. 
The assignment of $Z_K$-charges to chiral 
multiplets is of great importance in explaining 
the observed hierarchical pattern of 
quark-lepton masses.

The mass hierarchy of up-type quarks is a direct 
result of the horizontal discrete symmetry. 
On the other hand, for down-type quarks there appears 
a mixing between $D^c$ and $D'^c(=g^c)$. 
Due to the maximal mixing mass pattern of down-type 
quarks is different from that of up-type quarks. 
The mass hierarchy obtained here is 
\begin{eqnarray}
    & & m_u = O(v_u\,x^{\alpha +\beta +\gamma +\delta }), 
                                           \quad 
        m_c = O(v_u\,x^{\gamma +\delta }), \quad 
        m_t = O(v_u),    \\ 
    & & m_d = O(v_d\,x^{\alpha +\beta +\gamma +\delta }), 
                                           \quad 
        m_s = O(v_d\,x^{\beta +\gamma +\delta }), \quad
        m_b = O(v_d\,x^{-\alpha +\beta +\delta }). 
\end{eqnarray}
These results are consistent with observations 
under the parametrization $\alpha =w$, $\beta =2.5w$, 
$\gamma =2w$, $\delta =1.5w$ 
and $x^w =\lambda \sim 0.22$. 
Further we obtain a phenomenologically viable 
CKM matrix. 
In lepton sector $L$-$L'(=H_d)$ mixing occurs. 
Hence, mass pattern of charged leptons is also 
changed from that of up-type quarks under 
a large mixing. 
The obtained mass hierachy for leptons is 
\begin{equation}
     m_e = O(v_d\,x^{\alpha +\beta +\gamma +\delta 
                           +\xi }),            \quad 
     m_{\mu } = O(v_d\,x^{\alpha +\gamma +\delta }), 
                                               \quad
     m_{\tau } = O(v_d\,x^{\gamma }). 
\end{equation}
The CKM matrix in lepton sector amounts to a 
unit matrix irrespectively of the magnitude of 
$L$-$L'(=H_d)$ mixing. 
This is because both $L$ and $L'(=H_d)$ are 
$SU(2)_L$-doublets. 
Therefore, lepton flavor violating processes 
are extremely suppressed. 
Seesaw mechanism is at work 
for neutrinos. 
For large $s$ $(s \geq 8)$ light neutrino masses 
are 
\begin{eqnarray}
     m_{\nu e} & = & \frac {v_u^2}{M_S} 
       O(x^{k+1-2e+2(\alpha + \gamma + \xi )}), \nonumber \\ 
     m_{\nu \mu } & = & \frac {v_u^2}{M_S} 
       O(x^{k+1-2e+2(\alpha + \gamma )}),              \\ 
     m_{\nu \tau } & = & \frac {v_u^2}{M_S} 
       O(x^{k+1-2e+2\gamma }).                  \nonumber  
\end{eqnarray}
These masses seem to be too small compared with 
those expected from solar neutrino and atmospheric 
neutrino data. 
In the present framework the proton lifetime is 
$10^{33 \sim 35}$yr, 
which is long enough to be consistent with 
experimental data. 
The suppression of the dimension-five operators 
occurs because of the superheavy mass of 
the mediating particle for certain processes 
and because of the extremely small couplings for 
the other processes. 
On the other hand, we are not succesfull in 
achieving the unification of 
gauge couplings at the string scale. 
Nevertheless, it is suggestive that 
the obtained numerical value 
$\alpha_{string}^{-1} \sim 14$ corresponds nearly 
to the self-dual point $g_{string} = 1$ 
with respect to $S$-duality (strong/weak duality).

Both in $D^c$-$D'^c(=g^c)$ and $L$-$L'(=H_d)$ mixings 
the mass differences between heavy states 
and light states are extremely large in 
order of magnitudes. 
This implies that these mixings do not 
practically bring about flavor-changing 
neutral current processes. 
In addition, flavor-changing neutral current 
processes via superparticle exchanges 
at loop level are also suppressed enough 
to be consistent with experimental data, 
provided that the soft SUSY breaking parameters 
are universal at the string scale. 
More explicitly, the most stringent 
experimental bound on the mass difference of 
squarks $\tilde d$ and $\tilde s$ is derived 
from the $K^0$-$\bar K^0$ mixing. 
As pointed out in section 4, 
$SU(2)_L$-singlet components of down-type 
quarks $d$, $s$ and $b$ are nearly 
$O(1)D^c_1 + O(1)D^c_2$, $D'^c_1$ and $D'^c_3$, 
respectively. 
Although $D^c$ and $D'^c$ are indistinguishable 
from each other under the standard model 
gauge group, 
$D^c$ and $D'^c$ reside in ${\bf (6^*, 2)}$ 
and ${\bf (15, 1)}$ of $SU(6) \times SU(2)_R$, 
respectively. 
Further, $D^c$ and $D'^c$ reside in 
${\bf (4^*, 1, 2)}$ and ${\bf (6, 1, 1)}$ of 
$SU(4)_{PS} \times SU(2)_L \times SU(2)_R$, 
respectively. 
Therefore, gauge interactions cause soft SUSY 
breaking masses of $\tilde d^c_R$ and 
$\tilde s^c_R$ to evolve differently through 
radiative corrections in the energy region 
ranging from $M_S$ to $M_S\,x^k$. 
However, in the present model this energy 
range is rather nallow. 
In fact, $x^k$ is about $10^{-0.8}$ in a 
typical example. 
Consequently, it can be shown that the 
difference $\delta m^2 = m^2(\tilde d^c_R) - 
m^2(\tilde s^c_R)$ remains small at low energies. 
Let us estimate numerically the difference 
$\delta m^2$ in a typical example. 
When we assume $\delta m^2 (M_S)= 0$, 
the difference at the scale $M_S\,x$ 
$(= \langle S_0 \rangle)$ becomes 
\begin{equation}
 \delta m^2 (M_S\,x) \simeq -0.016 \times M_A^2 
\end{equation}
through the RG evolution, 
where $M_A$ is an averaged gaugino mass. 
Subsequently, the RG evolution from $M_S\,x$ 
to $M_S\,x^k$ $(= \langle N^c_0 \rangle)$ 
leads to 
\begin{equation}
 \delta m^2 (M_S\,x^k) - \delta m^2 (M_S\,x) 
     \simeq 0.008 \times M_A^2. 
\end{equation}
Combining these two results, we obtain 
\begin{equation}
 \delta m^2 (M_S\,x^k) \simeq -0.008 \times M_A^2. 
\end{equation}
Since Yukawa couplings of down-type quarks 
are tiny in case of $\tan \beta \sim 1$, 
the contributions of Yukawa interactions to 
$\delta m^2 (m_{3/2})$ are small compared with 
$\delta m^2 (M_S\,x^k)$. 
It follows that $\delta m^2 (m_{3/2})$ 
$\simeq $ $\delta m^2 (M_S\,x^k)$, 
which is consistent with a bound on 
$\delta m^2 (m_{3/2})$ given 
in Ref.\cite{KKB}.

Although we did not deal with CP-violation, 
there are two possibilities of introducing 
the CP-phase in the present framework. 
One possibility is that the CP-phase comes from 
complex VEVs of moduli fields. 
In this case the coefficients of the terms 
in the string-scale superpotential are complex 
in general. 
Another possibility is the case that 
the coefficients in the superpotential 
are all real but VEVs 
$\langle S_0 \rangle$, $\langle {\overline S} \rangle$, 
$\langle N_0^c \rangle$ and 
$\langle {\overline N^c} \rangle$, 
are complex. 
When we take the relative phase of 
$\langle N_0^c \rangle \langle {\overline N^c} \rangle$ 
to $\langle S_0 \rangle \langle {\overline S} \rangle$ 
into account, 
there appears CP-violating phase in the model.

\newpage 

\appendix 
\section*{Appendix A}
\setcounter{equation}{0}
\renewcommand{\theequation}{A.\arabic{equation}}

In this appendix we show that 
the minimization of the scalar potential yields 
tree-level breaking of the gauge symmetry 
under an appropriate condition on soft SUSY 
breaking parameters. 
In minimal supergravity model the soft SUSY 
breaking terms are given by
\cite{soft}
\begin{eqnarray}
      {\cal L}_{soft} & = & \int d^4\theta \,
              \Phi ^{\dag } \left[ 
                m_{3/2}\,\theta ^2 B 
              + m_{3/2}\,{\overline \theta ^2} B^* 
              - m_{3/2}^2 \,\theta ^2 
                         {\overline \theta ^2} C 
              \right] \exp (2gV) \Phi 
                                        \nonumber   \\
       & & \hphantom{QQQQ}
            - \left[ \int d^2\theta \,m_{3/2}\,
               \theta ^2 A\, W  + {\rm h.\,c.} \right]. 
\end{eqnarray}
Here $m_{3/2}$ is supposed to be $O(1{\rm TeV})$. 
The universal soft SUSY breaking parameters 
$A$, $B$ and $C$ are generally zero or order unity. 
Although $A$ and $B$ are generally complex numbers, 
$C$ is a real one. 
This type of ${\cal L}_{soft}$ leads to 
the scalar potential 
\begin{eqnarray}
     V & = & \sum_i \left| \frac {\partial W}
                      {\partial \phi _i} \right|^2 
             + \,m_{3/2}\bigl( A \,W + A^* \,W^*\bigr) 
                                       \nonumber   \\
       &   & + \,m_{3/2}\sum_i \Bigl( B\,\phi _i 
               \frac {\partial W}{\partial \phi _i}
                 + B^* \,\phi _i^* 
              \frac {\partial W^*}{\partial \phi _i^*}
                        \Bigr)      \nonumber   \\
       &   & + \,m_{3/2}^2 (C + |B|^2) \sum_i |\phi _i|^2 
             + (D{\rm -term}), 
\end{eqnarray}
where $\phi _i$ is a scalar component of 
the chiral superfield $\Phi _i$. 
In the above expression it is assumed 
that the terms of higher powers of $1/M_{pl}$ 
are negligibly small. 
We will shortly show that this assumption is 
justified.

For illustration, we take one set of vector-like 
multiplet $\Phi $ and ${\overline \Phi }$, 
whose scalar components are denoted as $\phi $ and 
${\overline \phi }$, respectively. 
Let us consider the case that 
the nonrenormalizable interaction 
\begin{equation}
    W = \lambda M_S^{3-2n}(\Phi {\overline \Phi })^n 
\end{equation}
is compatible with the discrete symmetry, 
where $\lambda $ is a positive $O(1)$ constant 
and $n$ is a large positive integer. 
In a typical example (\ref{eqn:nsk}) 
we put $n=sk=50$. 
For simplicity we denote dimensinless 
quantities $V/M_S^4$ and $\phi /M_S$ by 
the same letters as the original $V$ and $\phi $. 
Thus 
\begin{eqnarray}
    V & = & n^2\lambda ^2 \left\{ \left| \phi ^{n-1}
                         {\overline \phi }^n\right| ^2
               + \left| \phi ^n
                    {\overline \phi }^{n-1}\right|^2 
                         \right\}        \nonumber   \\
       &  & + \rho \lambda \left\{ (A+2nB)
                       (\phi {\overline \phi})^n 
            + (A^*+2nB^*)(\phi ^*{\overline \phi}^*)^n 
                    \right\}              \nonumber  \\
       &  & + \rho ^2(C+|B|^2)(|\phi |^2 
                         + |{\overline \phi }|^2) 
          + (D{\rm -term}) 
\end{eqnarray}
with $\rho = m_{3/2}/M_S$. 
Minimization of $V$ leads to the $D$-flat direction 
\begin{equation}
      |\langle \phi \rangle | = 
       |\langle {\overline \phi }\rangle | = x. 
\end{equation}
Writing the phase factor of VEVs explicitly as 
\begin{equation}
      \langle \phi \rangle 
       \langle {\overline \phi }\rangle  
                       = x^2 e^{i\theta }, 
\end{equation}
we have the scalar potential 
\begin{equation}
     V = 2n^2\lambda ^2 x^{4n-2} 
       + 2\rho \,|A+2nB|\,\lambda x^{2n} \cos \delta 
       + 2\rho ^2(C+|B|^2)x^2, 
\end{equation}
where $\delta = n\theta + \arg (A+2nB)$. 
From the stationary condition 
$\partial V/\partial \delta = 0$ and $x \geq 0$ 
the phase $\theta $ is determined as $\cos \delta =-1$. 
Therefore, the dependence of $V$ on $x$ is given by 
\begin{equation}
   V  =  2 \left[ n \lambda x^{2n-1} 
           - \rho \,\left| B + \frac {A}{2n} \right| 
                                    \,x \right]^2 
           - 2 \rho ^2 \left[ \left| B + 
                        \frac {A}{2n} \right|^2 
              - C - |B|^2 \right] \,x^2. 
\end{equation}
Consequently, if the inequality 
\begin{equation}
     \left| B+\frac {A}{2n} \right|^2  >  C + |B|^2 
\end{equation}
holds, 
$V$ is minimized at a nonzero value of $x$, 
namely at $x \sim \rho ^{1/(2n-2)}$. 
If $C \leq 0$, the above inequality is satisfied,
for example, in the case $|\arg(A\,B^*)| < \pi /2$ 
even for large $n$. 
It is worth emphasizing that the soft SUSY breaking 
mass parameter $(C+|B|^2)$ is not necessarily 
negative. 
If only the above inequality is satisfied, 
the gauge symmtry is spontaneously broken 
at tree level. 
It is not necessary for us to rely on the radiative 
symmetry breaking mechanism. 
In this paper the exponent $n$ is taken to be 
rather large. 
The larger $n$ implies the larger VEV 
$|\langle \phi \rangle | = M_S x$. 
The large value of $| \langle \phi \rangle |$ 
is consistent with the tree-level symmtry breaking.

In supergravity theory with canonical K\"ahler 
potential the supersymmetric term of the 
scalar potential is expressed as 
\begin{equation}
   V = e^{K/M^2}\left[ \,\sum_i 
      \left| \frac{\partial W}{\partial \phi _i} 
      + \frac{\phi_i^*}{M^2} W \right|^2 
      - \frac{3}{M^2}|W|^2 \right] + (D{\rm -term}) 
\end{equation}
with $M = M_{pl}/\sqrt {8\pi } \gsim M_S$. 
In the present model we get 
\begin{eqnarray}
     \left| \langle \frac{\partial W}
               {\partial \phi } \rangle \right|  
           & \simeq & n\lambda x^{2n-1}M_S^2, 
                                 \nonumber    \\
     \left| \langle \frac{\phi^*}{M^2} 
                W \rangle \right|  
           & \simeq & \lambda x^{2n+1} 
                    \bigl(\frac{M_S}{M}\bigr)^2 M_S^2, 
                                              \\
     \left| \langle \frac{1}{M} W \rangle \right| 
           & \simeq & \lambda x^{2n} 
                    \bigl(\frac{M_S}{M}\bigr) M_S^2. 
                                 \nonumber 
\end{eqnarray} 
Since $n$ is large and $x < 1$, $M_S/M \lsim 1$, 
$V$ is dominated by $\partial W/\partial \phi $. 
The overall factor $\langle \exp(K/M^2) \rangle $ 
is order unity. 
Therefore, the above analysis is relevant to 
the issue of the symmetry breaking.

\newpage 

\section*{Appendix B}
\setcounter{equation}{0}
\renewcommand{\theequation}{B.\arabic{equation}}

In this appendix we address to the issue of 
$R$-parity conservation within the present 
framework. 
It is shown that if eigenvalues of the mass 
matrix 
\begin{equation}
    \widehat{M}_{NS} = \left(
        \begin{array}{cc}
           x^{2k}N     &  x^{k+1}T^T  \\
           x^{k+1}T^T  &   x^2S  
        \end{array}
        \right)
\end{equation}
in $M_S$ units are sufficiently large compared 
with $m_{3/2}$, 
the scalar potential is minimized along 
the direction where $R$-parity is conserved. 
The mass matrix $\widehat{M}_{NS}$ is a submatrix 
of $\widehat{M}_{LNS}$ given in section 5 
and yields masses of $R$-parity odd and 
$G_{st}$-neutral superfields. 
It has already been found in section 5 that 
the above condition is satisfied for the solutions 
discussed in the text.

The superpotential can be separated as 
\begin{equation}
    W = W_1 + W_2, 
\end{equation}
where $W_1$ is a function only of $R$-parity 
even fields $S_0$, ${\overline S}$, $N^c_0$ 
and ${\overline N^c}$, 
while each term of $W_2$ contains $R$-parity 
odd fields $\Phi _i = S_j,\,N^c_j$ 
$(i=1,\cdots, 6\,;\ j=1,\,2,\,3)$. 
In the same manner as the notations in appendix A, 
we now use dimensionless quatities in $M_S$ units. 
Due to the $Z_K \times Z_2$ symmetry 
the explicit form of $W_1$ is given by 
\begin{equation}
     W_1 = \sum_{r=0}^s c_r 
            (S_0 {\overline S})^{(s-r)k}
            (N^c_0 {\overline N^c})^r, 
\end{equation}
where $c_r$ are $O(1)$ constants in $M_S$ units. 
This superpotential satisfies a relation 
\begin{equation}
     W_1 = \frac{1}{2sk} \left[ 
            S_0 \frac{\partial W_1}{\partial S_0} 
              + {\overline S} \frac{\partial W_1}
                  {\partial {\overline S}} \right] 
          + \frac{1}{2s} \left[ 
            N^c_0 \frac{\partial W_1}{\partial N^c_0} 
              + {\overline N^c} \frac{\partial W_1}
                  {\partial {\overline N^c}} \right]. 
\end{equation}
$W_2$ is a even function of $\Phi _i$. 
Consequently, the scalar potential is of the form 
\begin{equation}
    V = V_1 + V_2 
\end{equation}
with 
\begin{eqnarray}
     V_1 & = & \left| \frac{\partial W}{\partial S_0}
           + \rho (B+ \frac{A}{2sk})^* S_0^* \right|^2 
        + \left| (S_0 \rightarrow {\overline S}) \right|^2 
                                   \nonumber   \\
    & & \hphantom{QQQ} 
        + \left| \frac{\partial W}{\partial N^c_0}
           + \rho (B+ \frac{A}{2s})^* N^{c*}_0 \right|^2 
        + \left| (N^c_0 \rightarrow {\overline N^c}) 
              \right|^2            \nonumber  \\
    & & \hphantom{QQQ} 
         -\rho ^2 \biggl( \left| B + \frac{A}{2sk}\right|^2 
             - C - |B|^2 \biggr)
             \biggl( |S_0|^2 + |{\overline S}|^2 \biggr) 
                                   \nonumber  \\
    & & \hphantom{QQQ}
         -\rho ^2 \biggl( \left| B + \frac{A}{2s}\right|^2 
             - C - |B|^2 \biggr)
             \biggl( |N^c_0|^2 + |{\overline N^c}|^2 \biggr), 
                                              \\
     V_2 & = & \sum_{i=1}^6 \left| 
                \frac{\partial W_2}{\partial \phi _i}
                + \rho B^* \phi _i^* \right|^2 
                + \rho (A\,\tilde{W_2} + A^*\,\tilde{W_2}^*) 
                + \rho ^2 C \sum _{i=1}^6 |\phi _i|^2. 
\end{eqnarray}
Here $\phi _i$'s $(i=1, \cdots, 6)$ 
represent scalar components of 
$S_j$ and $N^c_j$ $(j=1,\,2,\,3)$ 
and $\tilde{W_2}$ is defined by 
\begin{equation}
    \tilde{W_2} = W_2 - \frac{1}{2sk} \left[
            S_0 \frac{\partial W_2}{\partial S_0}
          + {\overline S} \frac{\partial W_2}
                  {\partial {\overline S}} \right] 
             - \frac{1}{2s} \left[
            N^c_0 \frac{\partial W_2}{\partial N^c_0}
          + {\overline N^c} \frac{\partial W_2}
                  {\partial {\overline N^c}} \right]. 
\end{equation}
Scalar components of $S_0$, ${\overline S}$, 
$N^c_0$ and ${\overline N^c}$ are denoted by 
the same letters as the superfields themselves. 
As discussed in appendix A, 
under the assumption 
\begin{equation}
    \left| B+\frac{A}{2sk} \right|^2,\ 
    \left| B+\frac{A}{2s} \right|^2 
                            > C+|B|^2, 
\end{equation}
$S_0$, ${\overline S}$, 
$N^c_0$ and ${\overline N^c}$ develop nonzero 
VEVs and then the gauge symmetry is spontaneously 
broken at tree level. 
The stationary condition is satisfied at nonzero 
values of $S_0$, ${\overline S}$, $N^c_0$ and 
${\overline N^c}$ and vanishing 
$\langle \phi _i \rangle $. 
At this stationary point we get a negative 
value of the scalar potential 
\begin{equation}
     V = V_1 = -O(\rho ^2 \langle S_0 \rangle ^2). 
\end{equation}
The question here is whether this point is 
the absolute minimum or not.

Let us suppose that some of $\phi _i$ develop 
nonzero VEVs at the absolute minimum point. 
For such $\phi _i$, if 
\begin{equation}
    \left| \langle \frac{\partial W_2}
        {\partial \phi _i} \rangle \right| 
        \gg \rho |\langle \phi _i \rangle |, 
\end{equation}
then $V_2$ is dominated as 
\begin{equation}
    V_2 \simeq \sum_i \left| 
          \langle \frac{\partial W_2}
           {\partial \phi _i} \rangle \right|^2 
        \gg \rho ^2 \sum_i 
                 |\langle \phi _i \rangle |^2 
\end{equation}
and lifts up the scalar potential $V$. 
It follows that this point can not be the 
absolute minimun. 
Therefore, the relation 
\begin{equation}
    \left| \langle \frac{\partial W_2}
        {\partial \phi _i} \rangle \right| 
        \lsim  \rho |\langle \phi _i \rangle | 
\label{eqn:upb}
\end{equation}
should be satisfied for all $i$. 
On the other hand, the mass matrix of 
$\phi _i(\Phi _i)$ is given by 
\begin{equation}
   \langle \frac {\partial ^2 W_2}
      {\partial \phi _i \partial \phi _j} \rangle 
       = \left( \widehat{M}_{NS} \right)_{ij}.
\end{equation}
This matrix yields masses of $R$-parity odd and 
$G_{st}$-neutral superfields, 
which are assumed to be sufficiently larger than 
$\rho = m_{3/2}/M_S$. 
Namely, when we introduce a unitary matrix 
$\widehat{U}_{NS}$ which diagonalizes 
$\widehat{M}_{NS}$, 
this assumption is expressed as 
\begin{equation}
      \sum_{j,k} \left( \widehat{U}_{NS}^{-1} 
         \right)_{ij} 
          \langle \frac {\partial ^2 W_2}
        {\partial \phi _j \partial \phi _k} \rangle 
                 \left( \widehat{U}_{NS} 
              \right)_{ki} 
            \gg \rho 
\label{eqn:lwb}
\end{equation}
for all $i$. 
Although we have six unknown parameters 
$\langle \phi _i\rangle $, 
there are twelve constraints 
(\ref{eqn:upb}) and (\ref{eqn:lwb}) 
on $\langle \phi _i\rangle $ 
in which the orders of magnitude are quite 
different. 
Since we have too much constraints on 
$\langle \phi _i\rangle $, 
in generic case there are no consistent solutions 
except for $\langle \phi _i\rangle = 0$ for all $i$. 
Consequently the absolute minimum of $V$ is 
achieved at $\langle \phi _i\rangle = 0$. 
This means that $R$-parity is conserved.


\newpage 

\begin{center}
{\large {\bf Table I  \\ }} 
\vspace {5mm} 
\begin{tabular}{|c|cc|ccc|} \hline 
\phantom{\Bigg(} & $\overline{\Phi }$ &  $\Phi _0$  
       &  $\Phi _1$  &  $\Phi _2$  &  $\Phi _3$  \\
                                              \hline 
\phantom{\Bigg(} $({\bf 15, 1})$ \quad &  $(\overline{a}, +)$  
             &  $(a_0, +)$ 
                      &  $(a_1, -)$  &  $(a_2, -)$  
                          & $(a_3, -)$  \\
\phantom{\Bigg(} $({\bf 6^*, 2})$ \quad &  $(\overline{b}, +)$  
             &  $(b_0, +)$ 
                      &  $(b_1, -)$  &  $(b_2, -)$  
                          & $(b_3, -)$  \\
                                              \hline 
\end{tabular} 
\end{center}

\vspace{2cm}

\begin{center}

{\large {\bf Table II  \\ }} 
\vspace {5mm} 
\begin{tabular}{|c|c|c|} \hline 
\phantom{\Bigg(}  $R$-parity  &  Matter fields 
       & \quad $X$ (mass scale:\ $m=O(M_S\,x^X)$\,)  \\
                                              \hline 
     & \vphantom{\Big(} $Q_0$, $L_0$, 
           ${\overline Q}$, ${\overline L}$,   &  1  \\
     &  $\frac{1}{\sqrt 2}(S_0-{\overline S})$  &   \\
                                           \cline{2-3} 
     & \vphantom{\Big(} $U^c_0$, $E^c_0$, 
           ${\overline U^c}$, ${\overline E^c}$,  
                                           &  $k$  \\
     &  $\frac{1}{\sqrt 2}(N^c_0-{\overline N^c})$ &  \\
                                           \cline{2-3} 
     & \vphantom{\Big(} 
          $\frac{1}{\sqrt 2}(S_0+{\overline S})$  
                          &  $2sk-2$   \\  \cline{2-3} 
 $+$  & \vphantom{\Big(} 
          $\frac{1}{\sqrt 2}(N^c_0+{\overline N^c})$ 
                          &  $2sk-2k$  \\  \cline{2-3} 
     & \vphantom{\Big(} $H_{u0}$, $H_{d0}$, 
           ${\overline H_u}$, ${\overline H_d}$, 
            &   $2sk-4k+2e-1$,    \\
     &      &   $2sk-2e+1$        \\   \cline{2-3} 
     &      &   $2k-2e-3$,        \\
     &  $g_0$, $g^c_0$, $D^c_0$, ${\overline g}$, 
         ${\overline g^c}$, ${\overline D^c}$     
            &   $2sk-3k+2e-2$,    \\
     &      &   $2sk-k-1-\xi $    \\
                                        \hline  
     & \vphantom{\Big(} $H_{ui}$, $\tilde{L}'_i$   
         &  $k+\beta +2\delta $, \ 
         $k+\delta $, \ 
         $k $                 \\  
                                      \cline{2-3} 
     & \vphantom{\Big(} $D'_i$, $\tilde{D}'^c_i$  
         &  $k+\alpha +\gamma +\delta $, \ 
            $k+\beta +\delta -\alpha $, \ 
            $k$                      \\ 
                                      \cline{2-3} 
  $-$ &     &  $-k+\gamma +2e-1$,          \\ 
     &  $N^c_i$, $S_i$  
         &  $-k+\alpha +\gamma +2e-1$,  \\ 
     &   &  $-k+\beta +2\delta +2e-1$   \\  
                                      \cline{2-3} 
     &  \vphantom{\Big(} 
        $Q_i$, $U^c_i$, $\tilde{D}^c_i$, $\tilde{L}_i$, 
        $E^c_i$  &  $ > 2sk-2$                  \\
                                      \hline 
\end{tabular} 
\end{center}

\newpage 

\begin{center}
{\large Table Captions}  \\
\end{center}

{\bf Table I} \qquad 
The numbers $a_i$ and $b_i$ $(i = 0, 1, 2, 3)$ 
in the parentheses represent the $Z_K$-charges 
of chiral superfields $\Phi ({\bf 15, 1})$ and 
$\Phi ({\bf 6^*, 2})$, respectively. 
${\overline a}$ and ${\overline b}$ stand for those of 
mirror chiral superfields 
${\overline \Phi }({\bf 15^*, 1})$ and 
${\overline \Phi }({\bf 6, 2})$, respectively. 
Respective $Z_2$-charges ($R$-parity) of the superfields 
are also listed.

\vspace{1cm}

{\bf Table II} \qquad 
Particle spectra in the present model. 
The number $X$ stands for the exponent of $x$ 
for the mass scale $m = O(M_S\,x^X)$ of 
each superfield. 
Note that $x^{2sk-2} = m_{3/2}/M_S$ and $K=sk+1$. 
The parameters $\alpha $, $\beta $, $\gamma $ 
and $\delta $ are given in section 4. 
In this table $\tilde{D}^c_i$ and $\tilde{D}'^c_i$ 
($\tilde{L}_i$ and $\tilde{L}'_i$) stand for 
light and heavy eigenstates, respectively, 
which are derived via mixings between 
$D^c_i$ and $D'^c_i$ ($L_i$ and $L'_i$).

\vspace{3cm}

\begin{center}
{\large Figure Captions}  \\
\end{center}

{\bf Fig.1} \qquad 
The renormalization group flow of gauge couplings. 
The string scale, the soft SUSY breaking scale 
and the unified gauge coupling are taken as 
$M_S = 0.5 \times 10^{18}$GeV, 
$m_{3/2} = 200$GeV and 
$\alpha _{string}^{-1} =14.0$, respectively. 
The Yukawa coupling of the third generation 
is fixed to $f=1.7$.



\newpage 
\begin{thebibliography}{1}

\bibitem{Frog}
C. Froggatt and H. B. Nielsen, Nucl. Phys. 
                          {\bf B147} (1979) 277. 

\bibitem{Bound}
M. Leurer, Y. Nir and N. Seiberg, Nucl. Phys. 
                          {\bf B398}  (1993) 319. \\
L. Ib\'a\~nez and G. G. Ross, Phys. Lett. {\bf B332} 
                                      (1994) 100. \\
P. Bin\'etruy and P. Ramond, Phys. Lett. {\bf B350} 
                                      (1995) 49. \\
V. Jain and R. Schrock, Phys. Lett. {\bf B352} 
                                      (1995) 83. 

\bibitem{LBL}
L. J. Hall and L. Randall, Phys. Rev. Lett. 
                  {\bf 65} (1990) 2939. \\
M. Dine, R. Leigh and A. Kagan, Phys. Rev. 
                  {\bf D48} (1993) 4269. \\
Y. Nir and N. Seiberg, Phys. Lett. {\bf B309} 
                    (1993) 337. \\
P. Pouliot and N. Seiberg, Phys. Lett. {\bf B318}
                    (1993) 169. \\
M. Leurer, Y. Nir and N. Seiberg, Nucl. Phys. 
                   {\bf B420}  (1994) 468. \\
D. B. Kaplan and M. Schmaltz, Phys. Rev. 
                   {\bf D49} (1994) 3741. \\
L. J. Hall and H. Murayama, Phys. Rev. Lett. 
                   {\bf 75} (1995) 3985. \\
C. D. Carone, L. J. Hall and H. Murayama, 
        LBL-38047(1995), hep-ph/9512399; 
     LBL-38380(1996), UCB-PTH-96/06, hep-ph/9602364. 

\bibitem{Orbi}
T. Kobayashi, Phys. Lett. {\bf B354} (1995) 264; 
        {\it ibid.} {\bf B358} (1995) 253. 

\bibitem{SO10}
G. Anderson, S. Dimopoulos, L. Hall, S. Raby and 
       G. Starkman, Phys. Rev. {\bf D49} (1994) 3660. \\
K. S. Babu and R. N. Mohapatra, Phys. Rev. Lett. 
                      {\bf 74} (1995) 2418. 

\bibitem{Nambu}
Y. Nambu, preprint EFI 92-37. \\
P. Bin\'etruy and E. Dudas, Phys. Lett. {\bf B338} 
       (1994) 23; Nucl. Phys. {\bf B442} (1995) 21. \\
G. Kounnas, I. Pavel and F. Zwirner, Phys. Lett. 
                             {\bf B335} (1994) 403. \\
G. Kounnas, I. Pavel, G. Ridolfi and F. Zwirner, 
                Phys. Lett. {\bf B354} (1995) 322. \\
P. Bin\'etruy and E. Dudas, Nucl. Phys. {\bf B451} 
               (1995) 31. 

\bibitem{Infra}
M. Lanzagorta and G. G. Ross, Phys. Lett. {\bf B349} 
                                         (1995) 319. 

\bibitem{Aligned}
N. Haba, C. Hattori, M. Matsuda, T. Matsuoka 
             and D. Mochinaga, Prog. Theor. Phys. 
                             {\bf 94} (1995) 233. 

\bibitem{Gepner}
D. Gepner, Phys. Lett. {\bf 199B} (1987) 380; \ 
           Nucl. Phys. {\bf B296} (1988) 757. 

\bibitem{Disc}
C. A. L\"utkin and G. G. Ross, Phys. Lett. {\bf 214B} 
                                     (1988) 357. \\
C. Hattori, M. Matsuda, T. Matsuoka and H. Mino, 
             Prog. Theor. Phys. {\bf 82} (1989) 599. 

\bibitem{Majorana}
N. Haba, C. Hattori, M. Matsuda, T. Matsuoka 
          and D. Mochinaga, 
             Phys. Lett. {\bf B337} (1994) 63; \ 
             Prog. Theor. Phys. {\bf 92} (1994) 153. 

\bibitem{Tadpole}
N. Haba, C. Hattori, M. Matsuda, T. Matsuoka 
          and D. Mochinaga, 
             Prog. Theor. Phys. {\bf 95} (1996) 191. 

\bibitem{Pati}
J. C. Pati and A. Salam, Phys. Rev. {\bf D10} 
                 (1974) 275. 

\bibitem{Miu}
J. E. Kim and H. P. Nilles, Phys. Lett. {\bf 138B} 
                 (1984) 150. \\
G. F. Giudice and A. Masiero, Phys. Lett. {\bf B206} 
                 (1988) 480. \\
J. A. Casas and C. Mu\~noz, Phys. Lett. {\bf B306} 
                 (1993) 288. \\
J. E. Kim and H. P. Nilles, Mod. Phys. Lett. {\bf A9} 
                 (1994) 3575. 

\bibitem{Hisano}
J. Hisano, H. Murayama and T. Yanagida, 
         Phys. Rev. {\bf D49} (1994) 4966. 

\bibitem{Double}
G. K. Leontaris and J. D. Vergados, Phys. Lett. 
                 {\bf B258} (1991) 111. \\
E. Papageorgiu and S. Ranfone, Phys. Lett. {\bf B282} 
                 (1992) 89. 

\bibitem{Seesaw}
M. Gell-Mann, P. Ramond and S. Slansky, 
  in {\it Supergravity}, ed. D. Freedman et al. 
              (North-Holland, Amsterdam, 1979). \\
T. Yanagida, KEK Lectures, ed. O. Sawada et al. (1980) 
          912; Phys. Rev. {\bf D23} (1981) 196. \\
R. Mohapatra and S. Senjanovi\'c, Phys. Rev. Lett. 
         {\bf 44} (1980) 912. 

\bibitem{Solar}
R. Davis, Jr. {\it et al.}, in Proc. of the 21st 
International Cosmic Ray Conference, Adelaide, 
Australia, 1989, ed. R. J. Protheroe (Univ. of 
Adelaide Press, Adelaide, 1990), {\bf 12} 143. \\
K. S. Hirata {\it et al.}, Phys. Rev. Lett. {\bf 65} 
(1990) 1297; {\it ibid.} {\bf 66} (1990) 9.   \\
GALLEX collab., Phys. Lett. {\bf B285} 
                              (1992) 376; 390. \\
SAGE collab., Phys. Rev. Lett. {\bf 67} (1991) 3332; 
       Phys. Lett. {\bf B328} (1994) 234. 

\bibitem{Atmos}
K. S. Hirata {\it et al.}, Phys. Lett. {\bf B205} 
    (1988) 416; {\it ibid.} {\bf B280} (1992) 146. \\
Y. Fukuda {\it et al.}, Phys. Lett. {\bf B335} 
          (1994) 237. \\
D. Casper {\it et al.}, Phys. Rev. Lett. {\bf 66} 
          (1991) 2561.  \\
R. Becker-Szendy {\it et al.}, Phys. Rev. {\bf D46} 
          (1992) 3720.  

\bibitem{Cosmo}
E. L. Wright {\it et al.}, Astrophys. J. {\bf 396} 
             (1992) L13.   \\
M. Davis, F. J. Summers and D. Schlegal, Nature 
        {\bf 359} (1992) 393.  \\
A. N. Taylor and M. Rowan-Robinson, Nature 
        {\bf 359} (1992) 396.  \\
J. A. Holtzman and J. R. Primack, Astrophys. J. 
         {\bf 405} (1993) 428. 

\bibitem{Suema}
D. Suematsu, preprint (1996) KANAZAWA-96-05, 
                hep-ph/9604257. 

\bibitem{Pdecay}
N. Sakai and T. Yanagida, Nucl. Phys. 
                      {\bf B402} (1982) 533. \\
S. Weinberg, Phys. Rev. {\bf D26} (1982) 287. \\
J. Ellis, D. V. Nanopoulos and S. Rudaz, 
           Nucl. Phys. {\bf B202} (1982) 43. \\
P. Nath, A. H. Chamseddine and R. Arnowitt, 
           Phys. Rev. {\bf D32} (1985) 2348. 

\bibitem{Pdcal}
R. Arnowitt and P. Nath, Phys. Rev. Lett. 
             {\bf 69} (1992) 725. \\
P. Nath and R. Arnowitt, Phys. Lett. 
             {\bf B287} (1992) 89. \\
J. L. Lopez, D. V. Nanopoulos and H. Pois, 
             Phys. Rev. {\bf D47} (1993) 46.  \\
J. L. Lopez, D. V. Nanopoulos, H. Pois and 
      A. Zichichi, Phys. Lett. 
                      {\bf B299} (1993) 262.  \\
J. Hisano, H. Murayama and T. Yanagida, Nucl. Phys. 
             {\bf B402} (1993) 46. \\
R. Arnowitt and P. Nath, Phys. Rev. 
             {\bf D49} (1994) 1479. 

\bibitem{RGE}
M. E. Machacek and M. T. Vaughn, Nucl. Phys. 
                     {\bf B222} (1983) 83. \\
D. R. T. Jones, Phys. Rev. {\bf D25} (1982) 581. \\
D. R. T. Jones and L. Mezincescu, Phys. Lett. 
                     {\bf 136B} (1984) 242. \\
A. Parkes and P. West, Phys. Lett. {\bf 138B} 
                         (1984) 99. \\
P. West, Phys. Lett. {\bf 137B} (1984) 371. \\
D. R. T. Jones and L. Mezincescu, Phys. Lett. 
                      {\bf 138B} (1984) 293. \\
J. E. Bjorkman and D. R. T. Jones, Nucl.Phys. 
                      {\bf B259} (1985) 533. 

\bibitem{PDG}
Particle Data Group, Phys. Rev. {\bf D50} (1994) 1173.  

\bibitem{KKB}
F. Gabbiani and A. Masiero, Nucl. Phys. 
                        {\bf B322} (1989) 235. \\
J. S. Hagelin, S. Kelly and T. Tanaka, 
      Nucl. Phys. {\bf B415} (1994) 293. \\
S. Dimopoulos and D. Sutter, Nucl. Phys. 
                  {\bf B452} (1995) 496. 

\bibitem{soft}
L. J. Hall, J. Lykken and S. Weinberg, Phys. Rev. 
                      {\bf D27} (1983) 2359. \\
A. Sen, Phys. Rev. {\bf D30} (1984) 2608; 
            {\it ibid} {\bf D32} (1985) 411. \\
G. F. Giudice and E. Roulet, Phys. Lett. 
                      {\bf B315} (1993) 107. \\
Y. Kawamura, H. Murayama and M. Yamaguchi, 
              Phys. Rev. {\bf D51} (1995) 1337. 



\end{thebibliography}
\end{document}